\documentclass[twocolumn, english, reprint, longbibliography, superscriptaddress,
               breaklinks=true, showkeys, showpacs=false, nofootinbib,aps]{revtex4}

\usepackage[english]{babel}
\usepackage{graphicx,dcolumn,bm}
\usepackage{subcaption}   
\usepackage{color}
\usepackage[colorlinks=true,hyperindex,allcolors=blue]{hyperref}
\usepackage{amsfonts,ulem}
\usepackage{amssymb}
\usepackage{amsmath}
\usepackage{latexsym}
\usepackage{times,txfonts}
\usepackage{mathrsfs}
\usepackage{braket}
\usepackage{mathtools}
\usepackage{quantikz}
\usepackage{tikz}
\usetikzlibrary{arrows.meta, positioning, calc}
\usepackage{float}
\usepackage{epstopdf}
\usepackage{titlesec}
\usepackage{ragged2e}   

\makeatletter
\long\def\@makecaption#1#2{%
  \vskip\abovecaptionskip
  \small
  \justifying               
  #1\@addpunct{.}\ #2\par   
  \vskip\belowcaptionskip
}
\makeatother

\titleclass{\subsubsubsection}{straight}[\subsubsection]
\newcounter{subsubsubsection}
\renewcommand{\thesubsubsubsection}{\thesubsubsection.\arabic{subsubsubsection}}

\titleformat{\subsubsubsection}
  {\normalfont\normalsize\centering}  
  {\thesubsubsubsection}{1em}{}

\titlespacing*{\subsubsubsection}
  {0pt}{1.5ex plus 1ex minus .2ex}{1ex plus .2ex}

\usepackage{physics}
\newcommand{\aspas}[1]{``#1''}
\newcommand{\Hilb}{\ensuremath{\mathcal{H}}}
\newcommand{\Density}[1]{\ensuremath{\mathcal{D}(#1)}}

\usepackage[capitalise]{cleveref}

\usepackage{hyperref}
\hypersetup{
    colorlinks=true,
    linkcolor=blue,
    filecolor=blue,      
    urlcolor=blue,
    }
\makeatletter
\@ifundefined{textcolor}{}
{
\definecolor{BLACK}{gray}{0}
\definecolor{WHITE}{gray}{1}
\definecolor{RED}{rgb}{1,0,0}
\definecolor{GREEN}{rgb}{0,1,0}
\definecolor{BLUE}{rgb}{0,0,1}
\definecolor{CYAN}{cmyk}{1,0,0,0}
\definecolor{MAGENTA}{cmyk}{0,1,0,0}
\definecolor{YELLOW}{cmyk}{0,0,1,0}
}
\pdfoutput=1
\hypersetup{colorlinks=true,citecolor=blue,linkcolor=cyan,urlcolor=blue,filecolor= green, breaklinks=true}
\usepackage{url}
\usepackage{breakurl}
\usepackage{hyperref}
\makeatother
\newtheorem{Observation}{Observation}
\newtheorem{definition}{Definition}
\newtheorem{theorem}{Theorem}
\newtheorem{corollary}{Corollary}

\def\a{\hat a}
\def\adag{\hat a^\dag}
\def\q{\hat q}
\def\p{\hat p}
\def\cov{\text{Cov}}

\begin{document}

\title{An introductory review of the theory of continuous-variable quantum key distribution: Fundamentals, protocols, and security}

\author{Maron F. Anka\href{https://orcid.org/0000-0001-6720-4620}
{\includegraphics[scale=0.05]{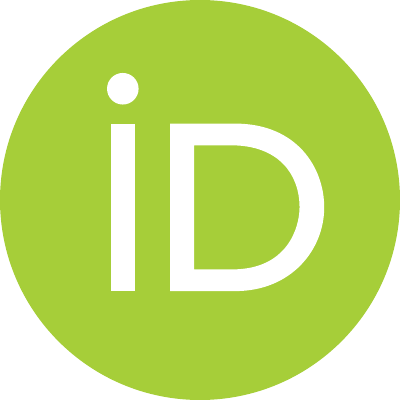}}}
\email{maron.anka@fbter.org.br}
\affiliation{QuIIN - Quantum Industrial Innovation, EMBRAPII CIMATEC Competence Center in Quantum Technologies, SENAI CIMATEC, Av. Orlando Gomes 1845, Salvador, BA, Brazil, CEP 41650-010.}

\author{John A. Mora Rodríguez \href{https://orcid.org/0000-0001-7625-1250}{\includegraphics[scale=0.05]{orcidid.pdf}}}
\email{john.rodriguez@fbter.org.br}
\affiliation{QuIIN - Quantum Industrial Innovation, EMBRAPII CIMATEC Competence Center in Quantum Technologies, SENAI CIMATEC, Av. Orlando Gomes 1845, Salvador, BA, Brazil, CEP 41650-010.}
\affiliation{Instituto de Matemática, Estatística e Computação Científica, Universidade Estadual de Campinas, CEP 13083-859, Campinas, Brazil}

\author{Douglas F. Pinto\href{https://orcid.org/0000-0002-3991-6935}{\includegraphics[scale=0.05]{orcidid.pdf}}}
\email{douglasfpinto@gmail.com}
\affiliation{QuIIN - Quantum Industrial Innovation, EMBRAPII CIMATEC Competence Center in Quantum Technologies, SENAI CIMATEC, Av. Orlando Gomes 1845, Salvador, BA, Brazil, CEP 41650-010.}

\author{Lucas Q. Galvão \href{https://orcid.org/0009-0001-2682-250X}{\includegraphics[scale=0.05]{orcidid.pdf}}}
\email{lqgalvao3@gmail.com}
\affiliation{QuIIN - Quantum Industrial Innovation, EMBRAPII CIMATEC Competence Center in Quantum Technologies, SENAI CIMATEC, Av. Orlando Gomes 1845, Salvador, BA, Brazil, CEP 41650-010.}

\author{Micael A. Dias \href{https://orcid.org/0000-0001-6394-9174}{\includegraphics[scale=0.05]{orcidid.pdf}}}
\email{mandi@dtu.dk}
\affiliation{Department of Electrical and Photonics Engineering, Technical University of Denmark, 2800  Lyngby, Denmark}
\affiliation{QuIIN - Quantum Industrial Innovation, EMBRAPII CIMATEC Competence Center in Quantum Technologies, SENAI CIMATEC, Av. Orlando Gomes 1845, Salvador, BA, Brazil, CEP 41650-010.}

\author{Alexandre B. Tacla}
\email{alexandre.tacla@fieb.org.br}
\affiliation{QuIIN - Quantum Industrial Innovation, EMBRAPII CIMATEC Competence Center in Quantum Technologies, SENAI CIMATEC, Av. Orlando Gomes 1845, Salvador, BA, Brazil, CEP 41650-010.}
 
\begin{abstract}
Continuous-variable quantum key distribution (CV-QKD) has emerged as a promising approach for secure quantum communication, offering advantages such as high key generation rates, compatibility with standard telecommunication infrastructure, and potential for integration on photonic chips. This review provides an accessible introduction to the theory of CV-QKD, aimed at researchers entering this rapidly developing field. We focus on fundamental concepts, key protocols, and security analysis essential for understanding CV-QKD systems, with a special emphasis on prepare-and-measure protocols using coherent states under asymptotic security conditions. We explain their equivalence to entanglement-based protocols and detail the security proof framework against collective attacks, encompassing both Gaussian and discrete modulation schemes. We also briefly address more advanced topics, including measurement-device-independent CV-QKD and finite-size security analysis. This work is motivated by Brazil's growing investment in quantum communication technologies. By presenting a clear learning path from basic concepts to advanced topics, this work aims to equip newcomers with the essential tools to engage with current research in CV-QKD, thereby supporting the training of a new generation of researchers in this strategic field.
\end{abstract}


\maketitle

\section{Introduction}
\label{introduction}

Quantum Key Distribution (QKD) is one of the most mature and promising applications of quantum information science. QKD enables secure communication over insecure channels, by allowing two parties to establish shared cryptographic keys protected by the fundamental principles of quantum mechanics. Since the pioneering work of Bennett and Brassard in 1984~\cite{Bennett1984}, the field has evolved from laboratory demonstrations to real-world implementations. Today, a variety of commercial QKD solutions are available from multiple vendors, several quantum networks have been demonstrated around the world~\cite{Cao2022,wei2022towards,liu2025road}, and satellite-based quantum communication systems~\cite{Liao2017,Yin2017, Ren2017,Bedington_2017, 
li2025microsatellite} have successfully established intercontinental quantum links~\cite{Liao_2018Satellite}. Most notably, the China Quantum Communication Network is the largest quantum network in the world to date, consisting of 145 fiber backbone nodes covering more than 10,000 km and linking 80 cities across 17 provinces. The network also includes six ground stations connected to the Jinan-1 quantum microsatellite~\cite{chen2025implementation}. This remarkable technological maturity has been primarily achieved through discrete-variable QKD (DV-QKD) systems, in which information is encoded in discrete degrees of freedom of a single photon, such as photon polarization. These systems have demonstrated robustness to noise and long-distance capabilities~\cite{liu2023experimental,Lasota2017, kish2024comparison}. Despite outstanding progress, practical challenges still limit widespread adoption of DV-QKD technologies. For instance, true single-photon sources are not widely available yet. On the receiver side, single-photon detectors are expensive and not standard telecom devices. Meanwhile, cheaper alternatives can only register signal detection, without distinguishing the exact number of photons in the incoming pulse. This limitation introduces a security loophole that an eavesdropper could exploit to obtain extra information from the transmitted signals \cite{Pirandola2020}.

In parallel, continuous-variable QKD (CV-QKD) has rapidly emerged as a promising complementary approach by exploiting the quadratures of optical fields as information carriers~\cite{Zhang2024}. In this way, CV-QKD offers distinct advantages, including higher key generation rates -- with recent demonstrations achieving Gbps rates for short distances ($\sim$10 km) and Mbps rates for longer distances ($\sim$100 km)~\cite{wang2025, hajomer2024}. Moreover, due to its similarity to classical optical communication systems, CV-QKD has a natural compatibility with standard commercially available telecommunication components and infrastructure, without requiring single-photon sources or detectors~\cite{Zhang2024, Diamanti2015}. This compatibility also enables full integration on photonic chips~\cite{hajomer2024b,hajomer2025chipbased}, opening possibilities for compact, scalable, and cost-effective quantum communication systems. Recent demonstrations have also shown the successful co-propagation of CV-QKD and classical data transmission channels over distances exceeding 120 km through optical fiber~\cite{hajomer2025coexistence}, showcasing CV-QKD as a potential ``plug-and-play" solution for metropolitan ($\lesssim$100 km) optical networks. Despite such promising advantages, CV-QKD systems also face inherent practical limitations. To ensure security, CV-QKD protocols must operate with very weak signals, which increases the system's sensitivity to imperfections and noise. This sensitivity limits viable communication distances and requires the use of complex classical post-processing procedures, including sophisticated error correction algorithms~\cite{Usenko2025, Pirandola2020}.

As quantum communication technologies mature worldwide, Brazil has also been increasingly investing in developing its own quantum communication capabilities, with major initiatives being developed~\cite{noticia_fapesp_RRQ_etc}, including metropolitan quantum networks in Recife, Rio de Janeiro~\cite{rederio}, and São Carlos, as well as the establishment in December 2023 of the EMBRAPII CIMATEC Competence Center in Quantum Technologies, called Quantum Industrial Innovation (QuIIN), which is developing Brazil's first point-to-point CV-QKD system, whose implementation is discussed in another article in this special issue of the Brazilian Journal of Physics.

The rapid development of quantum communication infrastructure in Brazil highlights the urgent need for training qualified researchers in this emerging field. Motivated by this growing demand, this review article offers an accessible introduction to the theory of CV-QKD, focusing on critical points that can be challenging for newcomers. Rather than attempting a comprehensive survey of the vast CV-QKD literature, we focus on discussing the fundamental concepts, key protocols, and security analysis that we believe are most essential for understanding this area. Specifically, we prioritize the discussion of prepare-and-measure (PM) protocols using coherent states under asymptotic security conditions. While the existing literature offers comprehensive reviews~\cite{djordjevic2019physical, Weedbrook2012, Pirandola2020, Usenko2025,Diamanti2015} and a more accessible tutorial~\cite{Laudenbach2018}, we found that certain fundamental concepts could benefit from additional accessible explanations, particularly for newcomers to the field. To further support the training of a new generation of Brazilian researchers specializing in QKD, a tutorial on quantum cryptography in Portuguese is presented in~\cite{sena2025}. Our goal here is to provide essential tools for those starting in CV-QKD, sharing the learning path that proved most effective for us. For advanced topics, we provide brief discussions with references to more detailed analyses, aiming to present currently relevant issues while directing attention to the fundamental topics necessary for their understanding. 

This article is organized as follows. In Section~\ref{preliminaryconcepts}, we introduce key concepts to understand the essential aspects of CV-QKD theory and give a review of classical and quantum information theory, specifically on entropy properties needed throughout the text, and which are often only briefly addressed in most existing articles. In Section~\ref{sec:cvqkd}, we present a brief overview of CV-QKD. In Sections~\ref{cvqkddescription} and~\ref{subsec1}, we provide a concise description of the PM CV-QKD protocol and its equivalence to the entanglement-based (EB) protocol. In Section~\ref{sec5}, we introduce the key points to understand and carry out security proofs in the asymptotic regime. Additionally, we present the trusted noise model, which has been widely used in the study of protocols' performance \cite{lodewyck2007quantum,fossier2009improvement,garcia2009continuous,usenko2016trusted}. Looking beyond PM protocols and asymptotic security, Sections~\ref{sec:MDI} and~\ref{sec6} introduce two advanced topics of significant current relevance in CV-QKD: measurement-device-independent CV-QKD (MDI-CV-QKD)  and  security analysis for the finite-length regime, respectively. These topics, while more complex than those in previous sections, represent critical frontiers in the field. We provide introductory treatments to serve as an entry point for readers interested in exploring these active research areas further.


\section{Preliminary concepts}
\label{preliminaryconcepts}

Before we dive into the details of CV-QKD theory, it is important to introduce some preliminary concepts from quantum optics and information theory that form the theoretical foundation of this field. The theory of CV-QKD brings together elements from continuous-variable quantum systems, classical information theory, and quantum information theory. In this section, we introduce some basic concepts that are essential for understanding CV-QKD protocols and their security analysis. We focus on the key concepts and mathematical tools most relevant for CV-QKD, providing the necessary background without attempting an exhaustive treatment of these vast fields.

\subsection{Continuous-variable quantum systems}
\label{sub:cv_systems}

We begin by giving a concise overview of basic concepts in quantum optics and key mathematical tools for describing continuous-variable quantum systems, which are defined in infinite-dimensional Hilbert spaces and are described by continuous-spectrum observables \cite{Braunstein2005, Navarrete2015, Weedbrook2012, hall2013quantum, serafini2023quantum}. For more detailed treatments of quantum optics and continuous-variable quantum information, we refer the reader to Refs.~\cite{Fabre2020, gerry2023introductory, scully1997quantum, mandel1995optical} and \cite{Braunstein2005,Weedbrook2012,adesso2014continuous,Navarrete2015,serafini2023quantum}, respectively. For readers who are not yet familiar with quantum information, we suggest the textbooks \cite{barnett2009quantum,nielsen2010quantum,wilde2013quantum}

\subsubsection{Quadrature operators}
\label{notations}

 The quantization of the free electromagnetic field provides the foundation for understanding continuous-variable quantum systems used in optical quantum communication. In the quantization procedure, the classical complex mode amplitudes are replaced by dimensionless non-Hermitian photon creation and annihilation operators ($\adag_k$ and $\a_k$, respectively), which satisfy the bosonic commutation relations
\begin{equation}
\begin{split}
&[\hat{a}_k,\hat{a}_{k'}] = [\hat{a}^{\dagger}_k,\hat{a}^{\dagger}_{k'}] = 0, \\
&[\hat{a}_k,\hat{a}_{k'}^{\dagger}] = \delta_{kk'},
\end{split}
\end{equation}
where the index $k$ labels each mode of the electromagnetic field, characterized by its wave vector $\vec k$, frequency $\omega_k$ and polarization $\vec \epsilon_k$~\cite{mandel1995optical, Fabre2020, Navarrete2015, gerry2023introductory}. This quantization process yields the Hamiltonian $\hat{H} = \sum_k \hbar \omega_k (\hat{a}_k^{\dagger} \hat{a}_k + 1/2)$ that is mathematically equivalent to a collection of independent quantum harmonic oscillators. 

The system's Hilbert space is the tensor product $\mathcal{H} = \bigotimes_{k=1}^m \mathcal{H}_k$, where $\mathcal{H}_k$ is the infinite-dimensional Hilbert space of mode $k$. Each mode is described by the continuous-spectrum quadrature operators $\q_k$ and $\p_k$, which act like the position and momentum operators of a quantum harmonic oscillator~\cite{gerry2023introductory}. In CV-QKD systems, the quadrature operators represent the measurable components of the optical field that are modulated to encode information. Following the usual convention in CV-QKD, we define the dimensionless quadrature operators in shot-noise units (SNU)~\cite{Braunstein2005, Weedbrook2012,Laudenbach2018}
\begin{align}
    \q_k &= \a_k + \adag_k, \label{eq:quadratureoperators_q} \\
    \p_k &= i(\adag_k - \a_k),\label{eq:quadratureoperators_p}
\end{align}
which, up to normalization factors, correspond to the real and imaginary parts of the annihilation operator. The quadrature operators are conjugated variables that satisfy the commutation relations 
\begin{equation}
\begin{split}
&[\hat{q}_k,\hat{q}_{k'}]=[\hat{p}_k,\hat{p}_{k'}] = 0 \\
&[\hat{q}_k,\hat{p}_{k'}]= 2i\delta_{k,k'}.
\end{split}
\end{equation}
Note that the definition of the quadrature operators, Eqs.(\ref{eq:quadratureoperators_q}) and (\ref{eq:quadratureoperators_p}), leads to commutation relations equivalent to setting $\hbar=2$ and, consequently, exhibits quantum fluctuations governed by the Heisenberg uncertainty relation
\begin{equation}
V(\hat{q}_k)V(\hat{p}_k) \geq 1, 
\label{uncertaintyrelation}
\end{equation}
where $V(\hat{A}) = \langle \hat{A}^2 \rangle - \langle \hat{A} \rangle^2$ denotes the variance of operator $\hat{A}$.

For convenience, all quadrature operators can be grouped into a single $2m$-component column vector $\boldsymbol{\hat{r}} = (\hat{q}_1, \hat{p}_1,..., \hat{q}_m, \hat{p}_m)^T$. In this compact notation, the canonical commutation relations of the quadrature operators can all be written as 
\begin{equation}
\label{eq:commutation_relations}
[\hat{r}_i, \hat{r}_j] = 2 i \Omega_{ij}, 
\end{equation}
where $\hat{r}_{2k-1} = \hat{q}_k$ and $\hat{r}_{2k} = \hat{p}_k$ for each mode $k = 1, 2,...,m$, and
\begin{equation}
\Omega = \bigoplus_{k=1}^m \omega = 
\begin{pmatrix}
\omega &  & \\
& \ddots & \\
& & \omega
\end{pmatrix},
\quad \text{with } \omega :=\begin{pmatrix}
0 & 1 \\
-1 & 0
\end{pmatrix},
\label{omega}
\end{equation}
is called the symplectic form \cite{Navarrete2015}. This representation is particularly useful, as we shall see later in this section, for describing Gaussian states and Gaussian operations, as it enables the compact handling of multimode systems~\cite{Weedbrook2012,serafini2023quantum}.

\subsubsection{Phase space representation and Gaussian states}

A quantum system is fully described by its density operator $\hat{\rho}$. An equivalent description is provided by quasiprobability distributions defined over the phase space, which are particularly useful to describe CV systems \cite{schleich2015quantum, barnett2009quantum}. Among various quasiprobability distributions, the Wigner function, defined by
\begin{equation}
W(q,p) \equiv \frac{1}{2\pi \hbar} \int_{-\infty}^{\infty} dy \,
\exp\!\left(-\tfrac{i}{\hbar} p y \right)
\left\langle q + \tfrac{1}{2}y \,\middle|\, \hat{\rho} \,\middle|\, q - \tfrac{1}{2}y \right\rangle,
\end{equation}
is particularly useful due to its close resemblance to classical probability distributions in both form and properties: it is unity normalized, $\int_{-\infty}^{\infty} dq \int_{-\infty}^{\infty} dp ~W(q,p) = 1$, and integration over one variable yields the correct marginal distribution for the other variable, i.e., $\int_{-\infty}^{\infty}  dp ~W(q,p) = \bra{q} \hat{\rho} \ket{q}$ and $\int_{-\infty}^{\infty} dq ~W(q,p) = \bra{p} \hat{\rho} \ket{p}$. The Wigner function is always real, but not in general positive, a feature that reveals the non-classical nature of quantum states. Nevertheless, it can be used to calculate the statistical moments of the quantum state.

Of special importance in CV quantum systems are Gaussian states, whose Wigner functions are Gaussian distributions in phase space~\cite{serafini2023quantum,schleich2015quantum}. The Wigner function of any $m$-mode Gaussian state can be written as
\begin{equation}
W(\boldsymbol{r},\Sigma) = \dfrac{1}{(2\pi)^m \sqrt{\det(\Sigma)}} e^{-\frac{1}{2}(\boldsymbol{r} - \boldsymbol{\bar{r}})^T \Sigma^{-1}(\boldsymbol{r} - \boldsymbol{\bar{r}})},
\end{equation}
where $\boldsymbol{\bar{r}}:= \langle \boldsymbol{\hat{r}} \rangle = tr(\boldsymbol{\hat{r}} \hat{\rho}) \in \mathbb{R}^{2m}$ is the vector of mean values, $\Sigma \geq 0$ is the $2m \times 2m$ positive-semidefinite symmetric covariance matrix (CM), whose elements are defined as $\Sigma_{ij} = \cov(\hat{r}_i,\hat{r}_j) = \frac{1}{2} \langle \{ \hat{r}_i - \langle \hat{r}_i \rangle, \hat{r}_j - \langle \hat{r}_j \rangle \} \rangle$, where $\{\cdot,\cdot\}$ is the anticommutator, and $\cov(\cdot,\cdot)$ is the covariance between the variables, representing their correlation. If the covariance term is zero for a given $\hat{r}_i$ and $\hat{r}_j$, then these modes are uncorrelated.

Hence, the state is completely characterized by its first two statistical moments. Note that the diagonal elements of the CM provide the variances of the quadrature operators, $\Sigma_{ii} = V(\hat{r}_i)$. Its general form can be written as
\begin{equation}
\small
\Sigma = 
\begin{pmatrix}
V(\hat{q}_1) & \cov(\hat{q}_1,\hat{p}_1) & \cov(\hat{q}_1,\hat{q}_2) & \cdots & \cov(\hat{q}_1,\hat{q}_m) & \cov(\hat{q}_1,\hat{p}_m) \\
& V(\hat{p}_1) & \cov(\hat{p}_1,\hat{q}_2) & \cdots & \cov(\hat{p}_1,\hat{q}_m) & \cov(\hat{p}_1,\hat{p}_m) \\
& & V(\hat{q}_2) & \cdots & \cov(\hat{q}_2,\hat{q}_m) & \cov(\hat{q}_2,\hat{p}_m)\\
& & & \ddots & \vdots & \vdots \\
& \text{symmetric} & & & V(\hat{q}_m) & \cov(\hat{q}_m,\hat{p}_m)\\
& & & & & V(\hat{p}_m)
\end{pmatrix}.
\label{cm}
\end{equation}

It is important to note that not all real, symmetric, positive matrices in the form above belong to the set of covariance matrices that represent a valid quantum state. Beyond the requirements for classical probability distributions, the covariance matrix of quantum states must satisfy additional constraints imposed by uncertainty relations. In this sense, Eq.(\ref{eq:commutation_relations}) provides necessary but not sufficient conditions, as it constrains only the quadrature variances (diagonal elements) but not the correlations between different quadratures (off-diagonal terms). The complete physical constraint ensuring that $\Sigma$ represents a valid quantum state $\hat{\rho}$ is given by the Robertson-Schrödinger uncertainty relation in matrix form \cite{serafini2023quantum}
\begin{equation}
\Sigma + i \Omega \geq 0,
\label{uncertainty}
\end{equation}
where the second term encodes the canonical commutation relation. This condition implies that the CM describing a valid quantum system must be positive-definite ($\Sigma > 0$). We refer the reader to Ref.\cite{Simon1994} for a full derivation of this relation. For example, in the single-mode case, we have det$(\Sigma + i \Omega) \geq 0$, which leads to the condition $V(\hat{q})V(\hat{p}) -\cov(\hat{q},\hat{p})^2 \geq 1$. This is precisely the Robertson-Schrödinger uncertainty relation, and we recover the standard Heisenberg uncertainty relation when $\cov(\hat{q},\hat{p}) = 0$.

The simplest Gaussian state is the vacuum state $\ket{0}$, with zero mean photon number $\hat{n}:= \hat{a}^{\dagger} \hat{a}$, $\langle \hat{n} \rangle = 0$. It belongs to a class of minimum uncertainty quantum states, saturating the equality in the Heisenberg principle $V(\hat{q})V(\hat{p}) = 1$ with equal variances $V(\hat{q}) = V(\hat{p}) = 1$. From the vacuum state, it is possible to obtain the most predominant classes of Gaussian pure states relevant to CV-QKD protocols, the coherent and squeezed states, via Gaussian operations (see Sec.~\ref{gaussop}). 

A coherent state is defined as the (right) eigenstate of the annihilation operator, $\hat{a} \ket{\alpha} = \alpha \ket{\alpha}$, with a complex eigenvalue $\alpha$. It can be written in the Fock basis as \cite{Navarrete2015}
\begin{equation}
\ket{\alpha} = e^{-|\alpha|^2/2} \sum_{n=0}^{\infty} \dfrac{\alpha^n}{\sqrt{n!}}\ket{n}.
\label{coherent}
\end{equation}
Coherent states form a non-orthogonal, overcomplete basis, and their overlap is given by $|\langle \beta| \alpha \rangle|^2 = e^{-|\beta-\alpha|^2}$ and average photon number $\langle \hat{n} \rangle = |\alpha|^2$. More generally, for a multimode Gaussian state, its mean photon number can be written in terms of its mean vector and CM as $\sum_{k=1}^m \langle\hat{n}_{k}\rangle = tr[\Sigma]/4 + \boldsymbol{\bar{r}}^2/4 - m/2$ \cite{Navarrete2015}. They also represent a minimum uncertainty state with equal variances $V(\hat{q}_k) = V(\hat{p}_k) = 1$.

Squeezed states, on the other hand, are not symmetric with respect to both quadratures. In fact, they are characterized by a reduction in the variance of one quadrature at the expense of an increase in the variance in the conjugate one. Nevertheless, this class of states still constitutes minimum uncertainty quantum states, with $V(\hat{q}) < 1$ and $V(\hat{p}) > 1$ or vice versa, depending on which quadrature the squeezing is applied. In the Fock basis, a single-mode squeezed state can be written as
\begin{equation}
\ket{s} = \sum_{n=0}^{\infty} \dfrac{1}{2^n n!} \sqrt{\dfrac{(2n)!}{\cosh{s}}} \tanh^n s \ket{2n},
\label{squeezed}
\end{equation}
where $s \in \mathbb{R}$ is the squeezing parameter, and the hyperbolic functions arise from the Bogoliubov transformation that defines the squeezing operator, which is discussed below. The state is squeezed in the $q$-direction for $s>0$ and in the $p$-direction for $s<0$.

\begin{figure}[t!]
    \centering
    \includegraphics[width=1\linewidth]{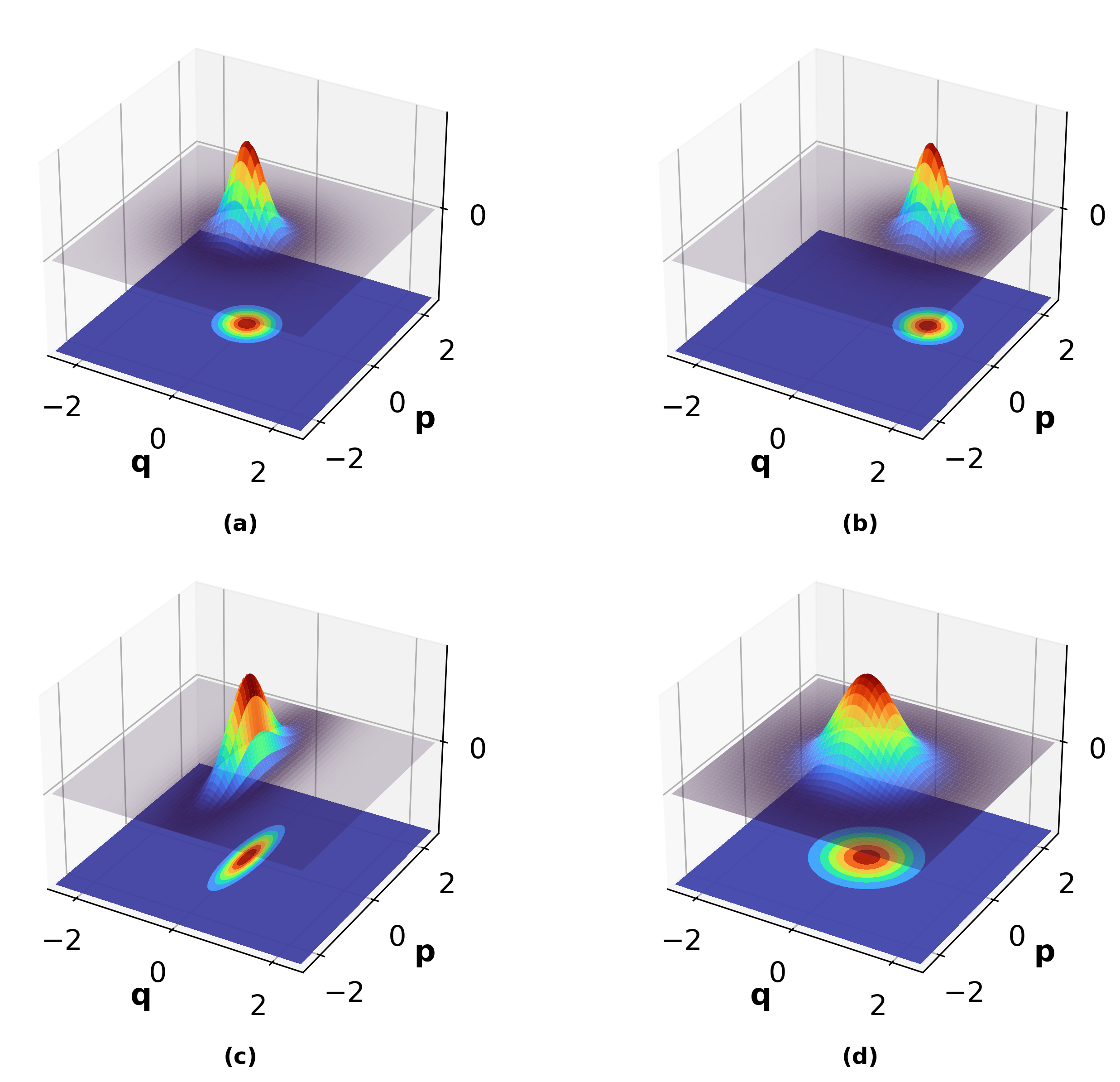}
    \caption{Representation of the Wigner functions for (a) vacuum, (b) coherent, (c) squeezed vacuum, and (d) thermal states in phase space.}
    \label{wignerstates}
\end{figure}

Another important example of a Gaussian state is the thermal state. By virtue of Williamson's theorem \cite{Williamson1936}, it is possible to show that any $m$-mode Gaussian state can be decomposed into $m$ uncorrelated thermal states \cite{Navarrete2015, Weedbrook2012} (see Sec.~\ref{symplecticformalism} for details). A thermal state is defined as the state that maximizes the von Neumann entropy $S = -tr(\hat{\rho}\log\hat{\rho})$, subject to a fixed energy $tr(\hat{n}\hat{\rho}) = \langle \hat{n} \rangle$, where $\langle \hat{n} \rangle > 0$. In the Fock basis, it can be written as
\begin{equation}
\hat{\rho}_{th} = \sum_{n=0}^{\infty}\dfrac{\langle \hat{n} \rangle^n}{(1 + \langle \hat{n} \rangle)^{n + 1}} \ket{n}\bra{n}.
\label{thermal}
\end{equation}
Since a maximally mixed state in an infinite-dimensional Hilbert space would correspond to an infinite number of excitations, i.e., $ \mathrm{tr}(\hat{n}\hat{\rho}) \to \infty$, such states are not physical. To ensure well-defined states, an energy constraint must be imposed, as was done above. This naturally leads to the definition of the thermal state, as given in Eq.(\ref{thermal}) \cite{Navarrete2015}. The CM representing the thermal state is described by $\Sigma = \text{diag}(V,V)$, where $V = 2 \langle\hat{n}\rangle + 1 > 1$. Thus, it is not a minimum uncertainty state \cite{Navarrete2015}. In Fig.~\ref{wignerstates}, we illustrate a schematic representation of the Wigner functions of the states discussed above in phase space.

Before we move on to the next section, we introduce an important state in CV-QKD that will be explored in later sections. The two-mode squeezed vacuum state (TMSVS)--also often referred to as an EPR\footnote{In the limit of $s \rightarrow \infty$, the TMSVS recovers the ideal EPR state \cite{Weedbrook2012}.} state--is an entangled two-mode Gaussian state. It can be written in the two-mode Fock basis as
\begin{equation}
\ket{\lambda} = \sqrt{1 - \lambda^2} \sum_{n=0}^{\infty} (-\lambda)^n \ket{n,n},
\label{tmsvs}
\end{equation}
where $\lambda = \tanh{s} \in [0,1]$, and $\ket{n,n}=\ket{n}\otimes\ket{n}$ represents a two-mode state with $n$ photons in each mode. Its CM is given by 
\begin{equation}
\Sigma = 
\begin{pmatrix}
u \mathbb{I} & \sqrt{u^2 - 1} \sigma_z \\
\sqrt{u^2 - 1} \sigma_z & u \mathbb{I}
\end{pmatrix},
\label{tmsvscm}
\end{equation}
where $\mathbb{I}=\text{diag}(1,1)$ and $u = \cosh{(2s)}$ quantifies the noise variance in the quadratures; the off-diagonal terms represent the correlations between the modes and $\sigma_z = \text{diag}(1,-1)$.

The CM allows one to easily evaluate the purity of a system. A Gaussian state is pure if $\sqrt{\det\Sigma} = 1$ and mixed otherwise \cite{adesso2014continuous}. In this sense, all states above are pure, except for thermal states. It is worth noting that the thermal state can be obtained by taking the partial trace over one of the modes of the TMSVS. In fact, the degree of entanglement of the global pure state is associated with the von Neumann entropy of the reduced state \cite{nielsen2010quantum}.

\subsubsection{Gaussian operations}
\label{gaussop}

Gaussian operations are defined as transformations that preserve the Gaussianity of quantum states -- that is, they map Gaussian states into Gaussian states. In the Heisenberg picture, these operations correspond to linear transformations, known as Bogoliubov transformations, that are generated by Gaussian unitary operators. They can be written in terms of the bosonic operators, but instead, we define them in terms of the quadrature operators \cite{Weedbrook2012,Navarrete2015}
\begin{equation} 
\boldsymbol{\hat{r}} \rightarrow \hat{U}^{\dagger} \boldsymbol{\hat{r}} \hat{U} = \mathcal{S} \boldsymbol{\hat{r}} + \boldsymbol{d},
\label{Bogoliubov}
\end{equation}
where $\boldsymbol{d} \in \mathbb{R}^{2m}$ and $\mathcal{S}$ is a $2m\times2m$ matrix. We can also describe the transformation acting directly on the mean vector and the CM of any Gaussian state: $\boldsymbol{\bar{r}} \rightarrow \mathcal{S} \boldsymbol{\bar{r}} + \boldsymbol{d}$ and $\Sigma \rightarrow \mathcal{S} \Sigma \mathcal{S}^T$. This transformation must preserve the canonical commutation relations of the quadrature operators. In this sense, the transformed quadrature operators must satisfy $[\hat{r}'_i,\hat{r}'_j] = 2 i \Omega_{ij}$, where $\boldsymbol{\hat{r}}' = \mathcal{S} \boldsymbol{\hat{r}} + \boldsymbol{d}$, with the $k$-th component $\hat{r}_k' = \sum_{l=1}^{2m} \mathcal{S}_{kl} \hat{r}_l + d_k$. The only non-vanishing term in the transformed commutation relation is $[\hat{r}'_i,\hat{r}'_j] = \sum_{ll'} \mathcal{S}_{il}\mathcal{S}_{jl'}[\hat{r}_{l},\hat{r}_{l'}] = 2i \sum_{ll'} \mathcal{S}_{il} \Omega_{ll'} \mathcal{S}_{jl'} = 2i (\mathcal{S} \Omega \mathcal{S}^T)_{ij}$, where we used Eq.(\ref{eq:commutation_relations}). Thus, in order to preserve the canonical commutation relation, the matrix $\mathcal{S}$ must satisfy
\begin{equation}
    \mathcal{S} \Omega  \mathcal{S}^T = \Omega,
\end{equation}
where $\Omega$ is defined in Eq.(\ref{omega}), implying that $\mathcal{S}$ must be a symplectic matrix \cite{Arvind1995}. It follows that the matrices $\mathcal{S}^{-1}$, $\mathcal{S}^T$, and $-\mathcal{S}$ are symplectic matrices as well. Using that $\Omega^T\Omega = \mathbb{I}$ and $\Omega^T=-\Omega$, we obtain $\mathcal{S}^{-1} = -\Omega \mathcal{S}^T\Omega$. The set of all $2m \times 2m$ real symplectic matrices constitutes a group denoted by Sp($2m,\mathbb{R}$). The transformation $\mathcal{S}$ corresponds to the action of an unitary $\hat{U}$ on the state $\hat{\rho}$. However, it only holds for unitary operations whose exponents are, at most, quadratic in regard of the bosonic operators $\hat{a}_k$ and $\hat{a}^{\dagger}_k$. Such Gaussian unitary transformations are called passive when they conserve the mean photon number of Gaussian states and active otherwise \cite{Navarrete2015}. A passive transformation is defined if and only if $\boldsymbol{d} = 0$ in Eq.\eqref{Bogoliubov} and $\mathcal{S}^T\mathcal{S} = \mathbb{I}_{2m\times2m}$, where the second restriction means that the symplectic transformation must be orthogonal, i.e., $\mathcal{S}^T = \mathcal{S}^{-1}$.

\subsubsubsection*{Examples of Gaussian operations}

The most relevant Gaussian unitaries for the CV-QKD formalism are shown in the following. The first one is the single-mode displacement operator, which is defined by the unitary operator $\hat{D}(\alpha) = e^{\alpha \hat{a}^{\dagger} - \alpha^* \hat{a}}$. Applying this operator to the vacuum state generates a coherent state, i.e. $\ket{\alpha} = \hat{D}(\alpha)\ket{0}$, as defined in Eq.(\ref{coherent}), which have the same properties as the vacuum state, except for its mean value that is shifted away from the origin, $\boldsymbol{d} = \{2\text{Re}(\alpha),2\text{Im}(\alpha)\}$. This operator generates a translation in phase space: the associated Bogoliubov transformation for a $m$-mode system has the form of Eq.(\ref{Bogoliubov}) with $\mathcal{S}_{\hat{D}(\alpha)} = \mathbb{I}_{2m \times 2m}$, implying that the CM is the same for the vacuum and coherent states. For a single mode, $\Sigma = \mathbb{I}$, with zero (non-zero) displacement for the vacuum (coherent) state. 

The second transformation is the single-mode squeezing operator $\hat{S}(s) = e^{s(\hat{a}^2 - \hat{a}^{{\dagger}2})/2}$, where $s\in [0,\infty)$ is the squeezing parameter. The effect of this transformation is to decrease the variance of one quadrature at the expense of increasing the other one. In the Heisenberg picture, the Bogoliubov transformation in terms of the bosonic operator is $\hat{a} \rightarrow \hat{S}^{\dagger}(s)  \hat{a} \hat{S}(s) = \hat{a}\cosh{s} - \hat{a}^{\dagger} \sinh{s}$, and $\boldsymbol{\hat{r}} \rightarrow \mathcal{S}_{\hat{S}(s)}\boldsymbol{\hat{r}}$ for the quadrature operators, with
\begin{equation}
\mathcal{S}_{\hat{S}(s)} =
\begin{pmatrix}
e^{-s} & 0\\
0 & e^s
\end{pmatrix}.
\end{equation}
The action of this operator on a vacuum state, $\hat{S}(s)\ket{0} = \ket{s}$, generates the so-called squeezed vacuum state, described by Eq.(\ref{squeezed}). The associated CM is $\Sigma = \text{diag}(e^{-2s},e^{2s})$.

Similarly to the single-mode squeezing operator, we can define the two-mode squeezing operator as $\hat{S}_2(s) = e^{s(\hat{a}_1 \hat{a}_2 - \hat{a}^{{\dagger}}_1\hat{a}^{{\dagger}}_2)}$. The Bogoliubov transformation associated with this operator is similar to the previous single-mode case: $\hat{a}_i \rightarrow S_2^{\dagger} (s) \hat{a}_i S_2(s) = \hat{a}_i\cosh{s} - \hat{a}^{\dagger}_j \sinh{s}$, with $\{i,j\} = \{1,2\}$ for each mode, or $\boldsymbol{\hat{r}} \rightarrow \mathcal{S}_{\hat{S}_2(s)}\boldsymbol{\hat{r}}$. Its matrix form is given by

\begin{equation}
\mathcal{S}_{\hat{S}_2(s)} =
\begin{pmatrix}
\cosh{s} ~ \mathbb{I} & \sinh{s} ~ \sigma_z\\
\sinh{s} ~ \sigma_z & \cosh{s} ~ \mathbb{I}
\end{pmatrix}.
\end{equation}
The action of this operator on the (two-mode) vacuum state generates the TMSVS, $\mathcal{S}_{\hat{S}_2(s)}\ket{0,0} = \ket{\lambda}$, as in Eq.(\ref{tmsvs}) with a CM matrix of the form of Eq.(\ref{tmsvscm}). This state plays a pivotal role in CV-QKD, as we shall see in Sec.~\ref{sec5}.

Another transformation that plays a fundamental role in CV-QKD is the beam splitter, which appears in theoretical models of Gaussian quantum channels \cite{Laudenbach2018} and coherent detectors \cite{usenko2016trusted,laudenbach2019analysis}. The beam splitter operator is given by $\hat{BS} = e^{\theta (\hat{a}^{\dagger}_1 \hat{a}_2 - \hat{a}_1 \hat{a}^{\dagger}_2)}$, where $\theta ~ \in [0,\pi/2]$ defines the transmissivity of the beam splitter $T = \cos^2{\theta} \in [0,1]$. The associated Bogoliubov transformation is $\hat{a}_1 \rightarrow \sqrt{T} \hat{a}_1 + \sqrt{1 - T} \hat{a}_2$ and $\hat{a}_2 \rightarrow -\sqrt{1 - T} \hat{a}_1 + \sqrt{T} \hat{a}_2$ or, in terms of the quadrature operators, $\boldsymbol{\hat{r}} \rightarrow \mathcal{S}_{\hat{BS}} \boldsymbol{\hat{r}}$, where

\begin{equation}
\mathcal{S}_{\hat{BS}} = 
\begin{pmatrix}
\sqrt{T} ~ \mathbb{I} & \sqrt{1 - T} ~ \mathbb{I}\\
-\sqrt{1 - T} ~ \mathbb{I} & \sqrt{T} ~ \mathbb{I}
\end{pmatrix}.
\label{beamspliter}
\end{equation}

The last example is the rotation transformation. The unitary operator that describes the rotation is given by $\hat{S}_{R} = e^{-i\theta \hat{a}^{\dagger}\hat{a}}$, where $\theta$ is the rotation angle. The associated Bogoliubov transformation for the annihilation operation can be written as $\hat{a} \rightarrow e^{i\theta} \hat{a}$, while the symplectic map corresponds to $\boldsymbol{\hat{r}} \rightarrow \mathcal{S}_{\hat{S}_{R}} \boldsymbol{\hat{r}}$, where

\begin{equation}
\mathcal{S}_{\hat{S}_{R}} = 
\begin{pmatrix}
\cos{\theta} & \sin{\theta} \\
-\sin{\theta} & \cos{\theta}
\end{pmatrix}.
\label{rotation}
\end{equation}

\subsubsubsection*{A few key properties from symplectic formalism}
\label{symplecticformalism}

The symplectic formalism provides a compact and rigorous mathematical framework for expressing the key properties of quantum states, particularly $m$-mode Gaussian states. In particular, an important result is the Williamson's theorem \cite{Williamson1936}, which states that for any $2m\times2m$ real positive-definite matrix can be diagonalized by a symplectic transformation \cite{serafini2023quantum}. In the case of the CM, there is a symplectic transformation $\mathcal{S}$ that brings $\Sigma$ to its diagonal form

\begin{equation}
\Sigma = \mathcal{S} \Sigma^{\oplus} \mathcal{S}^{T}, ~~~~~ \Sigma^{\oplus} = \bigoplus_{k=1}^m \nu_k \mathbb{I},
\label{williamson}
\end{equation}
where $\Sigma^{\oplus}$ is the diagonal form of $\Sigma$ in the so-called Williamson form, and $\{\nu_k\}_{k=1}^m$ are the symplectic eigenvalues of $\Sigma$. They can be obtained as the eigenvalues of the matrix $|i \Omega \Sigma|$, where $|X|=\sqrt{X^{\dagger}X}$. Physically, $\Sigma^{\oplus}$ can be seen as the CM of $m$ independent modes in a thermal state with mean photon numbers $\{ \langle \hat{n}_k \rangle = (\nu_k -1)/2 \}_{k=1,2,...,m}$, while $\mathcal{S}$ corresponds to a Gaussian unitary transformation. This theorem allows one to write the uncertainty principle, Eq.(\ref{uncertainty}), more compactly as $\Sigma > 0$ and $\Sigma^{\oplus} \geq \mathbb{I}$ \cite{Weedbrook2012}, or simply by stating that the symplectic eigenvalues must satisfy $\nu_k \geq 1$, for $k = 1, ..., m$, to represent a physical system. For instance, for a single-mode Gaussian state, the symplectic eigenvalue is given by $\nu_1 = \sqrt{\det \Sigma}$.

Two-mode Gaussian states are among the most important quantum states in CV quantum information, especially in CV-QKD \cite{Weedbrook2012}. They are widely used due to their analytical simplicity, including the study of entanglement in infinite-dimensional systems \cite{Navarrete2015,serafini2023quantum,adesso2014continuous}. The general form of the CM of a generic two-mode Gaussian state $\hat{\rho}_{AB}$, for modes $A$ and $B$\footnote{In order to introduce a more familiar convention used in the context of CV-QKD, we change the numerical notation previously used to denote the modes of different fields.}, can be written as
\begin{equation}
\Sigma_{AB} =
\begin{pmatrix}
\gamma_A & \gamma_{AB} \\
\gamma_{AB}^T & \gamma_B
\end{pmatrix},
\end{equation}
where $\gamma_A = \gamma_A^T$, $\gamma_B = \gamma_B^T$, and $\gamma_{AB}$ are $2\times2$ real matrices, which represent the quadrature variance of modes $A, B$, and their correlation, respectively. The general form of its symplectic eigenvalues is given by

\begin{equation}\label{eq:ev}
\nu_{1,2} = \sqrt{\frac{1}{2} \Bigg( \Delta \pm \sqrt{\Delta^2 - 4 ~ \Gamma} \Bigg)},
\end{equation}
where $\Delta = \det\gamma_A + \det\gamma_B + 2 \det\gamma_{AB}$, and $\Gamma = \det\Sigma_{AB}$. In this case, the uncertainty principle can be written as $\det\Sigma_{AB} \geq 1$ and $\Delta \leq 1 + \det\Sigma_{AB}$ \cite{Navarrete2015}.

A particular and important case of a two-mode Gaussian state CM is the so-called \textit{standard form}, which is written as

\begin{equation}
\Sigma_{AB} =
\begin{pmatrix}
a ~ \mathbb{I} & \gamma_{AB} \\
\gamma_{AB} & b ~ \mathbb{I}
\end{pmatrix}, ~~
\gamma_{AB} =
\begin{pmatrix}
c_1 & 0 \\
0 & c_2
\end{pmatrix},
\label{standardform}
\end{equation}
where $a, b, c_1, \text{and} ~ c_2 \in \mathbb{R}$. It is possible to show that the CM of any bipartite Gaussian state can be brought to this form through a local Gaussian unitary transformation \cite{Duan2000} (see Sec.~\ref{symmetrization section}). In particular, for $c_1 = -c_2 := c \geq 0$, $\gamma_{AB} = c ~\sigma_z$, the symplectic eigenvalues assumes the simple form 

\begin{equation}
\nu_{1,2} = \frac{\sqrt{(a + b)^2 - 4 c^2} \pm (b-a)}{2}.
\label{eq:gaussian ev}
\end{equation}
For more details on the symplectic formalism, we refer to Refs.\cite{Weedbrook2012,serafini2023quantum}

\subsubsection{Gaussian measurements}

\begin{definition}[Homodyne detection \cite{serafini2023quantum}]
Let the operator $\hat{x}_{\varphi}=\cos (\varphi) \hat{q} + \sin (\varphi) \hat{p}$, where $\hat{q}$ and $\hat{p}$ are the quadrature operators defined in Eqs.\eqref{eq:quadratureoperators_q} and \eqref{eq:quadratureoperators_p}, respectively. The homodyne detection scheme consists in the measurement of the rotated quadrature operator $\hat{x}_{\varphi}$, with outcome probability density
\begin{equation}
    p(x_{\varphi})=\langle x_{\varphi}|\hat{\rho}|x_{\varphi}\rangle, 
\end{equation}
where $\ket{x_{\varphi}}$ is an eigenvector of $\hat{x}_{\varphi}$ \cite{Weedbrook2012}.
\end{definition}

In the physical implementation of this measurement, the input mode, corresponding to the signal of interest, is mixed with a very intense local oscillator at a balanced beam splitter, followed by a detection of the intensity difference of the output modes. In a homodyne measurement, the local oscillator has the same frequency as the input signal, but a relative phase of $\varphi$, which determines which combination of the quadratures $\hat{q}$ and $\hat{p}$ is being measured.

\begin{definition}[Heterodyne detection \cite{serafini2023quantum}]
The heterodyne detection is described by the POVM $\{\pi^{-1}\ket{\alpha}\bra{\alpha}\}_{\alpha\in\mathbb{C}}$ [see Eq.(\ref{coherent})]. The outcomes are labelled by the coplex amplitude $\alpha$ and the probability density on a quantum state of a single mode $\hat{\rho}$ is given by 
\begin{equation}
    p(\alpha)=\frac{1}{\pi}\langle \alpha|\rho|\alpha\rangle.
\end{equation}
\end{definition}

The implementation of the heterodyne detection, which allows for the measurement of both quadratures simultaneously, is carried out similarly to the homodyne case, but with different frequencies between the input signal and the local oscillator. In CV-QKD, heterodyne detection is typically realized, however, through double homodyne detection (also known as an eight-port homodyne detector~\cite{ferraro2005}): a balanced beam splitter divides the signal into two paths, each followed by a homodyne detector measuring orthogonal quadratures. An important aspect of this setup is that one of the input ports of the beam splitter is left unoccupied, which effectively corresponds to a vacuum state entering the system \cite{Laudenbach2018}. This vacuum mode contributes an additional unit of shot noise to the measurement. As a result, while heterodyne detection enables the simultaneous measurement of both quadratures, it comes at the cost of introducing extra noise compared to homodyne detection, where only a single quadrature is measured at a time.

\subsection{Essentials of Classical and Quantum Information Theory}
\label{sub:information_theory}

The task of distributing binary sequences with secrecy is grounded in inherent quantum physical phenomena, such as non-orthogonal quantum states, the impossibility of a universal copy machine, the incompatibility of measurements (uncertainty principle), and so on \cite{Wootters1982, Bennett_1992, Horodecki2009, nielsen2010quantum}. Therefore, a theory for information transmission in quantum systems is essential to understand how much secrecy a quantum protocol can distribute (or generate). On the other hand, as will be discussed in Sec.~\ref{sec:V}, the resulting secret keys are classical random variables, and most of the practical part of a CV-QKD protocol take places in the classical domain through classical information processing. The goal of this section is to give the information-theoretical foundations of quantum key distribution. Since a detailed treatment of these topics is beyond the scope of this review, we focus on the essential concepts and main results. For comprehensive coverage, detailed derivations, and further discussion of the topics presented in this section, we refer interested readers to the following textbooks \cite{cover2006,polyanskiy2025,nielsen2010quantum,wilde2017b}.

    \subsubsection{Classical entropic quantities}\label{sec:classical-information-theory}

    The main object of classical information theory is the random variable and, consequently, its probability distribution. In his seminal paper \cite{shannon1948}, Claude E. Shannon laid down the basis for modern communication systems by addressing two basic questions about information systems: what is the most efficient representation of an information source, and how much information can be reliably transmitted over a noisy communication channel? These tasks are known as source and channel coding\footnote{It is also common to use source compressing instead of source coding.}, whose operational limits are given by entropic quantities, which are going to be briefly exposed throughout this section. 

    Consider a discrete random variable $X$ taking values from the alphabet $x\in\mathcal{X}$ and having a probability mass function $p(x)$. The entropy of $X$ is defined as
    \begin{equation}\label{eq:shannon-entropy}
        H(X) = -\sum_{x\in\mathcal{X}}p(x)\log(p(x)),
    \end{equation}
    \noindent with the logarithm taken in base $2$, which is standard in most of the literature, and gives a sense of information in units of \textit{bits} \footnote{When the natural logarithm is used, the entropy is given in units of $nats$.}. The entropy of a random variable as given in Eq.(\ref{eq:shannon-entropy}) is related to the notions of uncertainty and randomness of the random variable $X$, or, in a more operational meaning, as a bound to the compression rate for any reliable compression algorithm for an information source modeled by the random variable $X$ \cite{cover2006}. Notice that $0\leq H(X)\leq\log|\mathcal{X}|$ with equality on the left if $p(x)$ is degenerate and on the right if $p(x) = \frac{1}{|\mathcal{X}|}$.
    
    The definition of entropy can be generalized for multiple random variables in a natural manner, as well as for conditioned distributions. For the two random variables $X$ and $Y$ with joint probability distribution $p_{XY}(x,y)$, the joint entropy $H(X,Y)$ is given by,
    \begin{equation}\label{eq:joint-entropy}
        H(X,Y) = -\sum_{x\in\mathcal{X}}\sum_{y\in\mathcal{Y}}p(x,y)\log(p(x,y)).
    \end{equation}
    \noindent On the other hand, the entropy of the random variable $Y$ having knowledge of $X$ is the average entropy of $Y$ conditioned to the occurrence of the event $X=x$,
    \begin{align}\label{eq:conditioned-entropy}\nonumber
        H(Y|X) &= \sum_{x\in\mathcal{X}}p(x)H(Y|X=x)\\
                &= -\sum_{x\in\mathcal{X}}\sum_{y\in\mathcal{Y}}p(x,y)\log(p(y|x)).
    \end{align}

    Notice that in the three cases defined above, $\log p(x)$, $\log p(x,y)$ and $\log p(y|x)$ are functions of random variables, being random variables by definition. Therefore, Eqs.(\ref{eq:shannon-entropy}) to (\ref{eq:conditioned-entropy}) can be rewritten as the expectations of logarithmic functions\footnote{In concordance with traditional notation of classical information theory textbooks, we use $\mathbb{E}$ as the expectation operator for random variables.}
    \begin{align}
        H(X) &= -\mathbb{E}_X\log p(X),\\
        H(X,Y) &= -\mathbb{E}_{XY}\log p(X,Y),\\
        H(Y|X) &= -\mathbb{E}_{XY}\log p(Y|X).
    \end{align}

    The entropy of a random variable is a function of its distribution and quantifies the uncertainty associated with the random variable. In several practical problems, the probabilistic behavior of a physical system may not always be known, so an approximate distribution $q(x)$ is used instead of the true distribution $p(x)$. The informational cost of using $q$ instead of $p$ is given by the Kullback–Leibler (K-L) divergence, also known as relative entropy, given by
    \begin{equation}\label{eq:kl-divergence}
        D(p(x)||q(x)) = \sum_{x\in\mathcal{X}}p(x)\log\frac{p(x)}{q(x)}.
    \end{equation}

    The K-L divergence is not a true metric as it is not symmetric and does not satisfy the triangle inequality. However, it is non-negative and equals zero if and only if $p=q$. It is worth pointing out that in Eq.(\ref{eq:kl-divergence}) it is assumed the conventions: $0\log\left(\frac{0}{0}\right)=0$, $0\log\left(\frac{0}{q}\right)=0$ and $p\log\left(\frac{p}{0}\right)=\infty$.

    Based on the notion of informational divergence, the mutual information between two random variables $X$ and $Y$ with joint distribution $p(X,Y)$ is defined as the divergence between the joint distribution and the product of the marginals:
    \begin{align}
        I(X;Y) &= D(p(x,y)||p(x)p(y)),\\
        &=\sum_{x\in\mathcal{X}}\sum_{y\in\mathcal{Y}}p(x,y)\log\frac{p(x,y)}{p(x)p(y)}.
    \end{align}
    It quantifies the amount of classical correlation between the pair of random variables and equals zero if and only if $X$ and $Y$ are independent, i.e., $p(X,Y)=p(X)p(Y)$.

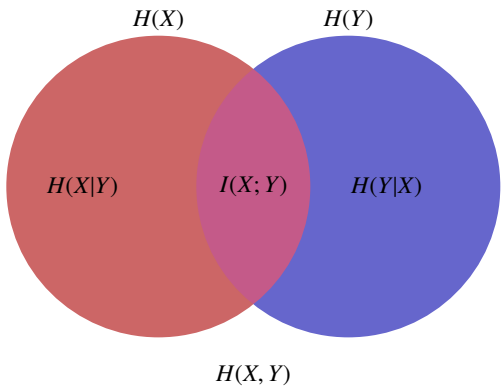
\begin{figure}
        \centering
        \begin{tikzpicture}
          \colorlet{colorA}{red!50}
          \colorlet{colorB}{blue!50}
          \colorlet{colorC}{green!50}
          \colorlet{blendcolor}{white!50} 
        
          \begin{scope}[blend group=soft light]
            \fill[colorA!80!black, name=x] (0,0) circle (2cm); 
            \fill[colorB!80!black, name=y] (2.5,0) circle (2cm); 
          \end{scope}
        
          \node at (0, 2.2) {$H(X)$};
          \node at (2.5,2.2) {$H(Y)$};
          \node at (1.25, 0) {$I(X;Y)$}; 
          \node at (-1, 0) {$H(X|Y)$}; 
          \node at (3, 0) {$H(Y|X)$}; 
          \node  at (1.25,-2.5) {$H(X,Y)$}; 
        \end{tikzpicture}
        \caption{Venn diagram for the relationship between entropies.}
        \label{fig:entropy-venn-diagramm}
    \end{figure}

    The entropic quantities defined above can be related according to the following identities:
    \begin{align}
        H(X,Y) &= H(X) + H(Y|X) = H(Y) + H(X|Y),\\
        I(X;Y) &= H(X) - H(X|Y) = H(Y) + H(Y|X), \label{mutualinfo} \\
               &= H(X) + H(Y) - H(X,Y), 
    \end{align}
    \noindent whose derivations can be found in \cite{cover2006}. It is also useful to use a graphical representation in the form of a Venn diagram, as in Fig.~\ref{fig:entropy-venn-diagramm}. The properties of these quantities, as well as the generalization for multiple random variables and the respective chain rules, can be found in the previously cited textbooks.

    The entropic quantities defined above can also be defined for continuous random variables. In general, extending the definition from discrete to continuous random variables requires replacing the summation with an integral. In this case, a random variable $X$ with a cumulative probability function (cdf) $F(x)=\operatorname{Pr}[X\leq x]$ is continuous if the function $F(x)$ is continuous. If the derivative $f(x) = F'(x)$ exists and $\int_{-\infty}^{\infty}f(x)=1$, then $f(x)$ is called the probability density function (pdf) of $X$.
    
    The entropy of a continuous random variable is called \textit{differential entropy} and is defined as
    \begin{equation}
        h(X) = -\int_Sf(x)\log f(x)dx,
    \end{equation}
    \noindent where $S$ is the support set of $X$, i.e., $S=\qty{x:f(x)>0}$. The definitions of joint, conditional entropy, and relative entropy are likewise using the pdf of the continuous random variables. For the relative entropy between densities $f$ and $g$, the functional $D(f||g)$ is bounded only if the support of $f$ is contained in the support of $g$. It is worth pointing out that, besides its similarities to the discrete case, the differential entropy is not invariant under general changes of variables and can take negative values.
    
    \subsubsection{Channels and capacity}

    As stated in the first paragraph of Sec. \ref{sec:classical-information-theory}, reliable information transmission lies at the heart of information theory. To provide a clearer view of the operational meanings of some quantities defined above, and also to provide useful results for the analysis of cryptographic protocols in the following sections, we must provide a formal definition of a communication channel and its informational capacity.

    A discrete channel maps input symbols from the alphabet $\mathcal{X}$ to output symbols in the alphabet $\mathcal{Y}$ according to a transition matrix $p(y|x)$ containing the probability of observing the output symbol $y$ when $x$ was input. The channel is said to be memoryless if the conditional probability of the input at time $i$ does not depend on previous channel usages, that is, $p(y_i|x_ix_{i-1}\cdots) = p(y_i|x_i) = p(y|x)$.

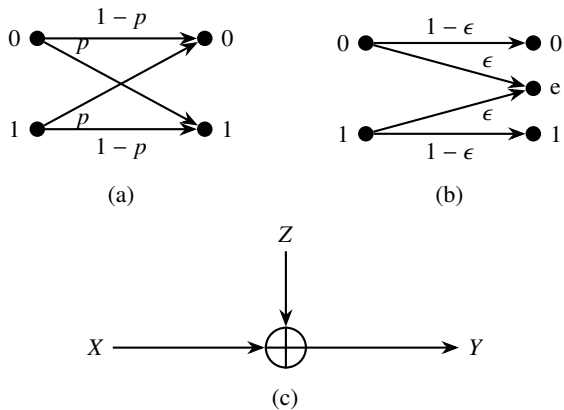
\begin{figure}[!tb]
        \centering
        \begin{subfigure}{.5\linewidth}\label{fig:bsc}
            \centering
            \begin{tikzpicture}[
                            node distance=1cm and 2cm,
                            box/.style={draw, circle, fill=black, minimum size=5pt, inner sep = 0, align=center},
                            >=Stealth, thick
                            ]
                \node[label=left:0] (x0) [box] {};
                \node[label=left:1] (x1) [box, below=of x0] {};
                \node[label=right:0] (y0) [box, right=of x0] {};
                \node[label=right:1] (y1) [box, right=of x1] {};
                \draw[->] (x0) -- node[above] {$1-p$} (y0);
                \draw[->] (x0) -- node[above, near start] {$p$} (y1);
                \draw[->] (x1) -- node[below] {$1-p$} (y1);
                \draw[->] (x1) -- node[below, near start] {$p$} (y0);
            \end{tikzpicture}
            \caption{}
        \end{subfigure}%
        \begin{subfigure}{.5\linewidth}\label{fig:bec}
            \centering
            \begin{tikzpicture}[
                            node distance=1cm and 2cm,
                            box/.style={draw, circle, fill=black, minimum size=5pt, inner sep = 0, align=center},
                            >=Stealth, thick
                            ]
                \node[label=left:0] (x0) [box] {};
                \node[label=left:1] (x1) [box, below=of x0] {};
                \node[label=right:0] (y0) [box, right=of x0] {};
                \node[label=right:1] (y1) [box, right=of x1] {};
                \node[label=right:e] (e) [box] at ($(y0)!0.5!(y1)$) {};
                \draw[->] (x0) -- node[above] {$1-\epsilon$} (y0);
                \draw[->] (x0) -- node[above, near end] {$\epsilon$} (e);
                \draw[->] (x1) -- node[below] {$1-\epsilon$} (y1);
                \draw[->] (x1) -- node[below, near end] {$\epsilon$} (e);
            \end{tikzpicture}
            \caption{}
        \end{subfigure}
        \begin{subfigure}{\linewidth}\label{fig:additive-channel}
            \centering
            \def\l{15}
            \begin{tikzpicture}[node distance=1cm and 2cm,>=Stealth, thick]
                \node[circle, draw, minimum size=\l pt] (O) at (0,0) {};
                \draw (O.south) -- (O.north);
                \draw (O.west) -- (O.east);
                \node[left=of O] (X) {$X$};
                \node[right=of O] (Y) {$Y$};
                \node[above=of O] (Z) {$Z$};
                \draw[->] (X) -- (O);
                \draw[->] (O) -- (Y);
                \draw[->] (Z) -- (O);
            \end{tikzpicture}
            \caption{}
            
        \end{subfigure}
        \caption{(a) Binary symmetric channel (BSC) with bit flipping probability $p$. (b) Binary erasure channel (BEC) with erasure probability $\varepsilon$. (c) Additive white Gaussian noise (AWGN) channel model.}
        \label{fig:bsc-bec}
    \end{figure}

    Some channels are useful for several information processing tasks, providing a good theoretical model for physical phenomena involving information transmission. Fig.~\ref{fig:bsc-bec} depicts the binary symmetric channel (BSC) and the binary erasure channel (BEC). The BSC models a channel that inverts the input bit with probability $p$, while the BEC erases the input bit with probability $\epsilon$. The noisy behavior of physical systems limits the amount of information that can be transmitted, but reliable communication can still be performed by adding controlled redundancy to the transmitted symbols. Take, for example, transmitting one bit of information through the BSC$(p)$ channel with $0<p<\frac12$. One simple strategy to protect the information bit is to use a repetition code of length $n$, which consists of transmitting $n$ copies of the information bit. For example, with $n=3$, the code performs the mappings $0\rightarrow000$ and $1\rightarrow111$. The receiver can use a majority rule to recover the original bit, which will be successful as long as just one error has happened. In this example, the code rate is $\frac13$, meaning that one information bit was transmitted using three code bits. Physically, this means the channel must be used three times to transmit one bit of information. The decoding error probability is then a function of the channel parameter $p$ and the code length, such that it is possible to make communication more robust by increasing the length $n$ of a repetition code, but at the cost of transmitting fewer information bits for each channel use.

    It is reasonable to question what the fundamental limit is for the number of bits that can be transmitted over a channel, regardless of the strategy used to protect information from noise\footnote{Coding theory is a field on its own, dedicated to the study the coding schemes and their properties. We recommend the interested reader to the following textbooks \cite{Blahu2003,Lin2024}.}. This number is known as the informational capacity of a communication channel, which is defined as
    \begin{equation}
        C = \max_{p(x)}I(X;Y),
    \end{equation}
    \noindent with the maximum taken over all input probability distributions on $\mathcal{X}$. One cornerstone of information theory is the fundamental result stating that reliable communication over a noisy channel is possible whenever the code rate is no greater than the channel capacity, $R\leq C$, which is known as the channel coding theorem \cite{shannon1948}. Some channels have closed expressions for their capacities. For the BSC$(p)$ and BEC($\epsilon$), their capacities are $1-H(p)$ and $1-\epsilon$, respectively, where we used $H(p) = -p\log p - (1-p)\log(1-p)$ as the binary entropy, and the capacities are reached with equiprobable input bits.

    Communication channels may also include continuous-valued alphabets at the output. The Gaussian channel is the most important of them, represented in Fig.~\ref{fig:bsc-bec}. It is an additive channel in the sense that at each channel use, the input value $X$ is added to a sample of  independent and identically distributed (i.i.d.) Gaussian random variables $Z\sim\mathcal{N}(0,N)$ independent of $X$, such that,
    \begin{equation}
        Y = X + Z.
    \end{equation}

    In practical problems, it is common to impose a power limitation on the inputs of the channel, such as an average power constraint. In this case, one considers that the input random variable satisfies $\mathbb{E}X^2\leq P$. The average power constrained Gaussian channel capacity is stated as
    \begin{equation}
        C = \max_{p(X)\,s.t.\,\mathbb{E}X^2\leq P} I(X;Y).
    \end{equation}

    As in the case for the BSC, the capacity of the Gaussian channel is known to be
    \begin{equation}
        C = \frac12\log(1+\frac{P}{N}),
    \end{equation}
    \noindent corresponding to the mutual information $I(X;Y)$ for $X\sim\mathcal{N}(0,P)$. That is, the capacity of a Gaussian channel with input average power constraint is met if the input symbols follow a zero-mean Gaussian distribution with variance $P$. The quantity $P/N$ is known as the signal-to-noise ratio (SNR) in its linear form and is a common parameter for the analysis of communication systems under noisy channels.

    \subsubsection{Quantum entropic quantities}
    
    In classical information theory, entropy is a function of a random variable's probability distribution that quantifies the uncertainty associated with that variable. Since uncertainty is an important concept in quantum theory, entropic measures occupy a place of great importance \cite{coles2017}. In quantum systems, the states are \aspas{probabilistic objects}, as their eigenspectra satisfy the properties of a probability distribution. Therefore, the entropy of a quantum system (a measure of uncertainty) is a function of the system's density operator, following the same spirit as Shannon's entropy for classical systems. This section introduces fundamental entropic functionals for quantum systems, as it was done for the classical case in the previous section. As before, we provide a concise overview of the key concepts. For comprehensive treatments and rigorous derivations of the material presented here, we refer readers to standard textbooks \cite{nielsen2010quantum,wilde2017b,watrous2018}.
        
    Let us start with the definition of the von Neumann entropy of a quantum state. Let $\vu\rho_A$ denote the density operator describing the quantum state of system $A$, which is defined on the Hilbert space $\Hilb_A$. The entropy $S(\vu\rho_A)$ of the state is defined as
    \begin{equation}\label{eq:entropy_vn}
        S(\vu\rho_A) = -\tr(\vu\rho_A\log\vu\rho_A).
    \end{equation}
        
    Since the density operator's spectral decomposition yields a probability distribution (its eigenvalues), the von Neumann entropy can be expressed as the Shannon entropy of this eigenvalue distribution, establishing a direct connection between quantum and classical information measures. Let $\lambda_A$ be eigenvalues and $\ket{\lambda_A}$ be eigenvectors of $\vu\rho_A$ such that they form the spectral decomposition $\vu\rho_A=\sum_A\lambda_A\op{\lambda_A}$. Then, the entropy of the quantum state is given by
    \begin{equation}\label{eq:entropia_vn_decompos}
        S(\vu\rho_A) = -\sum_A\lambda_A\log\lambda_A = H(\lambda),
    \end{equation}
    \noindent which is the Shannon entropy of the eigenvalues\footnote{The eigenvalues of a density operator correspond to a probability distribution since $\lambda_i\geq0$ for any $i$ and $\sum_i\lambda_i=1$.} of $\vu\rho_A$. 
    Analogously to the classical case, the joint entropy of a composite quantum system $AB$ with density operator  is defined as $\vu\rho_{AB}$,
    \begin{eqnarray}\label{eq:entro_conj}
    S(\vu\rho_{AB}) = -\tr(\vu\rho_{AB}\log\vu\rho_{AB}).
    \end{eqnarray}

    Some entropy relations of classical systems do not transfer directly to quantum systems, one of which is the case of states of joint systems. In the classical case, for example, if $X$ and $Y$ are two random variables, the inequality $H(X)\leq H(X,Y)$ is always true, since the addition of a system can only increase the uncertainty about the complete system. In the quantum case, the composite system can have uncertainty smaller than that of the parts. For example, consider the entangled Bell state $\ket{\Psi^+} = (\ket{00}+\ket{11})/\sqrt{2}$. As a pure bipartite state, it has zero entropy: $S(\ket{\Psi^+}) = 0$. However, the reduced density matrices of the individual subsystems are maximally mixed, each with entropy $S(\tr_A(\op{\Psi^+})) = S(\tr_B(\op{\Psi^+})) = 1$.

    This counterintuitive behavior with respect to the entropy of classical systems can be interpreted operationally from the definition of conditional entropy for quantum states. Let $\vu\rho_{AB}\in\Density{\Hilb_A\otimes\Hilb_B}$ be the state of a joint quantum system $AB$. The conditional quantum entropy $S(A|B)$ is defined as the difference between the joint entropy $S(\vu\rho_{AB})$ and the marginal entropy $S(B) = S(\vu\rho_B)$,
    \begin{equation}\label{eq:entr-cond}
    S(A|B) = S(A,B) - S(B).
    \end{equation}

    It is also possible to define a quantum version of the relative entropy, which is useful in developing various results in quantum information theory. Let $A$ be a quantum system and $\vu\rho,\,\vu\sigma\in\Density{\Hilb_A}$. The relative entropy between the states $\vu\rho$ and $\vu\sigma$ is defined as
    \begin{equation}\label{eq:relative_entropy}
    S(\vu\rho||\vu\sigma) = \tr(\vu\rho\log\vu\rho) - \tr(\vu\rho\log\vu\sigma).
    \end{equation}
    
    In Eq.(\ref{eq:relative_entropy}), for $S(\vu\rho||\vu\sigma)$ to assume finite values, the support of $\vu\rho$ must be contained in the support of $\vu\sigma$. If this condition is not met, the support of $\vu\rho$ will have a nontrivial intersection with the kernel of $\vu\sigma$, implying the non-separability of the states. As in the classical case, the quantum relative entropy seems to provide a distance measure between states, but it fails to be symmetric and satisfy the triangle inequality. However, it is possible to show that it is non-negative and that $S(\vu\rho||\vu\sigma)=$ only if $\vu\rho=\vu\sigma$.

    Another fundamental concept of quantum information theory (and of special interest in the analysis of CV-QKD protocols) is the notion of accessible information of a quantum system. It has no classical counterpart, and, in short, it is the maximal classical mutual information obtained in a scenario where the transmitter sends classical information by using an ensemble of quantum systems, and the receiver uses the optimal set of measurement operators. Formally, Alice (the transmitter) encodes the random variable $X$ by preparing quantum states according tothe ensemble $\qty{\vu\rho_x, p_x}$: when $X = x$ occurs with probability $p_x$, she sends state $\vu\rho_x$. Bob (the receiver) measures the received state using a set of positive operator-valued measurements (POVM) $\qty{E_Y}$, obtaining as measurement outcome the random variable $Y$. The alphabets of $X$ and $Y$ do not need to be the same, nor have the same size. In this sense, the classical mutual information between the transmitter and receiver variables is bounded above by
    \begin{equation}\label{eq:holevo-information}
        I(X;Y) \leq S\qty(\sum_xp_x\vu\rho_x) - \sum_xp_xS(\vu\rho_x) = \chi(X,Y),
    \end{equation}
    \noindent where the rightmost quantity is known as the Holevo bound~\cite{Holevo1973}.
    
    Before moving on to the next section, we briefly show the usefulness of the symplectic formalism to compute the von Neumann entropy for Gaussian states. In this context, using the relations from Eq.(\ref{williamson}), such states can be described as a tensor product of thermal states in the form of Eq.(\ref{thermal}), $\hat{\rho} = \bigotimes_{k=1}^{m} \hat{\rho}_{th_k}$, each with variance $V = 2 \langle\hat{n}_k\rangle + 1$. The resulting entropy is the sum of each thermal state composing the original Gaussian state, and it can be written as \cite{serafini2023quantum}
    \begin{equation}
    S_{vN} = \sum_{k=1}^m g(\nu_k),
    \label{eq:g function}
    \end{equation}
    where
    \begin{equation}
    g(x) := \Big(\dfrac{x + 1}{2}\Big) \log\Big(\dfrac{x + 1}{2}\Big) -  \Big(\dfrac{x - 1}{2}\Big) \log\Big(\dfrac{x - 1}{2}\Big)
    \label{eq:g function_}
    \end{equation}
    is the bosonic entropy function, which is positive and monotonically increasing for $x \geq 1$. A step-by-step derivation of this formula can be found in Ref.\cite{Demarie2018}.
    
    Since the von Neumann entropy is invariant under unitary transformations, the displacement operator does not affect the entropy. This means that the first moments (mean values) do not contribute to entropy calculations. Therefore, without loss of generality, Gaussian states in CV-QKD are typically assumed to have zero mean, with all relevant information contained in the covariance matrix \cite{Laudenbach2018}.

\section{Fundamentals of CV-QKD}
\label{sec:V}

\subsection{Brief historical overview}
\label{sec:cvqkd}

The early stages of CV-QKD, developed between 1999 and 2001, primarily relied on squeezed states as information carriers \cite{Ralph1999,Ralph2000,Hillery2000,Reid2000}. At that time, squeezing was regarded as a necessary non-classical resource for ensuring security, as coherent states were not yet sufficient \cite{Ralph2000}. The random choice of squeezed quadrature played a role analogous to the use of non-orthogonal bases in discrete-variable protocols such as the pioneering BB84~\cite{Bennett1984}. In these schemes, Alice encodes information in either the amplitude or phase quadrature of a squeezed state, while Bob measured the corresponding quadrature via homodyne detection \cite{Hillery2000}, and the security was guaranteed by the Heisenberg uncertainty relation, Eq.(\ref{uncertaintyrelation}). Although conceptually significant, these protocols were limited by the need of high levels of squeezing, sensitivity to channel loss, and the use of direct reconciliation. Moreover, their security analysis was restricted to individual attacks, lacking general proofs against collective or coherent strategies.

The first fully continuous CV-QKD protocol was proposed in 2001 by Cerf \textit{et al.} \cite{Cerf2001}: Alice encodes Gaussian-distributed symbols in the quadratures of a squeezed beam, while Bob measures either the $q$- or $p$-quadrature at random. This protocol enables secure shared information between the trusted parties, while potentially leaking information to an eavesdropper under the assumption of individual attack. In the same year, another squeezed beam protocol based on Gaussian modulation was proposed by Gottesman and Preskill \cite{Gottesman2001}, where it was the first to give a complete discussion of (discrete) error correction and privacy amplification, and it was also shown that the security against arbitrary attacks could be achieved using quantum error-correcting codes. By the end of 2001, Assche, Cardinal, and Cerf proposed the extension of the reconciliation and privacy amplification processes aimed for discrete-variables protocols to the continuous-variables case  \cite{Vanasche2004}\footnote{The paper was only published in 2004.}, which was a necessary step in order to provide more efficient CV-QKD protocols.

Following the squeezed-based protocols and the newly developed CV reconciliation and privacy amplification procedures, in 2002, the seminal protocol GG02 was introduced by Grosshans and Grangier \cite{Grosshans2002}. This marked a turning point in CV-QKD: it eliminated the need for squeezing by using coherent states with Gaussian modulation, and most modern protocols are built as extensions or refinements of the GG02. This is the first Gaussian modulated (GM) coherent state-based protocol with homodyne detection and security proven for individual attacks. The original proposal considered direct reconciliation (DR) of information, in which Alice serves as the reference for the post-processing stage, i.e., Bob corrects his data with respect to Alice's. For the protocol to remain secure, Bob must hold more information about Alice's sent states than Eve. In a purely lossy channel, once the channel transmittance $T$ falls to or below $50\%$, an eavesdropper (Eve) can potentially extract as much or more information than Bob. This threshold corresponds to a channel attenuation of 3 dB, which is the origin of the well-known 3 dB loss limit of the DR information process. Subsequently, the reverse reconciliation (RR) of information was introduced as an alternative to overcome this limitation, designating Bob as the reference \cite{grosshans2002reversereconciliationprotocolsquantum,Grosshans2003} and enabling security for individual attacks for arbitrary channel transmittance. In 2004, the No-switching protocol was proposed \cite{Weedbrook2004}, replacing homodyne detection with a double homodyne detection in GM coherent state protocols, thereby eliminating the need for random quadrature measurements and allowing for simultaneous measurement of both quadratures. It is worth noting that, in the CV-QKD literature, the terms heterodyne detection and double homodyne detection are often used interchangeably, even though they do not refer to the same physical process \cite{schleich2015quantum,muller2012quadrature}. For a pedagogical introduction to GM coherent state CV-QKD and noise modeling, we refer to Ref.\cite{Laudenbach2018}.

In 2006, the security level of CV-QKD protocols was extended from individual to collective attacks by Navascés et al. \cite{navascues2006optimality} and García-Patrón \cite{garcia2006unconditional}, who showed that Gaussian attacks are optimal among coherent attacks. In this context, the Gaussian extremality property \cite{wolf2006extremality} played an important role. In 2009, Renner and Cirac generalized the application of the quantum de Finetti theorem for infinite-dimensional systems, proving that it can be used to establish security against coherent (arbitrary) attacks in CV-QKD protocols, provided that they are secure against collective attacks \cite{renner2009finetti}. In the same year, García-Patrón and Cerf introduced a squeezed state protocol employing heterodyne detection under collective attacks, which was shown to outperform the previously mentioned GM protocols in the high-noise quantum channel regime \cite{garcia2009continuous}. This marked the reintroduction of protocols based on squeezed states in CV-QKD, not as a necessary condition for security but as a resource for improving metrics \cite{usenko2011squeezed}. For a review of squeezed state protocols, we refer the reader to Ref.\cite{nag2025continuous}.

The development and subsequent security analysis of CV-QKD encompass an enormous family of protocols, whose detailed description is beyond the scope of this work. However, we mention some examples to guide the interested reader: unidimensional protocol with coherent and squeezed states \cite{usenko2015unidimensional,usenko2018unidimensional}, security analysis in trusted noise model \cite{usenko2016trusted,laudenbach2019analysis}, proven security in free-space \cite{pirandola2021limits}, non-Gaussian operations-based protocols such as photon subtraction \cite{guo2017performance} and quantum scissors \cite{ghalaii2020long}, and complete elimination of information leakage using squeezed states \cite{jacobsen2018complete,winnel2021minimization}.

Even though GM CV-QKD protocols are optimum regarding their performance, they faced experimental issues, such as the precision limitation of the in-phase and quadrature (I/Q) modulators and the low post-processing efficiency \cite{bloch2006ldpc,leverrier2008multidimensional,leverrier2010continuous}. An alternative solution emerged in the discrete modulation approach. They were first proposed in 2002 to deal with the 3 dB limitation of the DR process by using binary encoding of coherent states \cite{silberhorn2002continuous}, assigning the bit value 0 (1) to a coherent state with positive (negative) displacement. Its security was proven against the individual \cite{silberhorn2002continuous} and collective attacks \cite{symul2007experimental,Heid2006,zhao2009asymptotic}, and a generalization was made for three states \cite{bradler2018security}. Other approaches were proposed for dealing with four \cite{leverrier2009} and eight states \cite{becir2012continuous} under the collective attacks assumption. The security proof for discrete modulated CV-QKD (DM CV-QKD) protocols is still an open question in the field. Some advances have been made in directions such as composable security under collective Gaussian attacks \cite{papanastasiou2021continuous} and general collective attacks using decoy states \cite{leverrier2011continuous}. More recently, Ghorai \textit{et al.} \cite{ghorai2019asymptotic} proposed an approach based on semidefinite programming (SDP), which enabled the establishment of a lower bound on the security against collective attacks for quadrature phase shift keying (QPSK) in the asymptotic limit. This result was generalized by Denys \textit{et al.} \cite{Denys2021} to arbitrary discrete modulations under collective attacks. Consequently, the investigation of amplitude-phase shift keying (APSK) and quadrature and amplitude modulation (QAM) CV-QKD protocols has gained attention \cite{almeida2021secret,almeida2023reconciliation}. Additionally, following the usefulness of convex optimization, Lin et al. \cite{Lin2019} developed another method that achieves better approximations than Ghorai’s method, as it is able to circumvent the use of Gaussian approximations and include post-processing steps in the analysis, albeit at the cost of increased computational complexity. The security analysis of GM and DM CV-QKD is discussed in more detail in Sec.~\ref{sec5}. 


\subsection{Protocol Description}
\label{cvqkddescription}

There are two standard formulations for implementing and analyzing QKD: the prepare-and-measure (PM) and the entanglement-based (EB) schemes. In the PM picture, Alice prepares quantum states of light and sends them to Bob through an insecure quantum channel; afterward, Bob performs measurements on the received signals. This is the most natural description for experimental implementations. In the EB picture, Alice and Bob share an entangled TMSVS; Alice measures one mode locally while the other is sent through the insecure quantum channel to Bob. The EB formulation is particularly convenient for security analyses, as it allows Eve’s interaction to be modeled as a purification of the shared state \cite{Laudenbach2018}. Although these operational descriptions differ, they are mathematically equivalent and yield identical statistics and key rates under the same physical assumptions \cite{Laudenbach2018,djordjevic2019physical}. Below, we introduce the step-by-step description of a PM CV-QKD protocol and then provide the relations between the PM and EB pictures.

A PM CV-QKD protocol can be divided into the following steps:
\begin{enumerate}\label{protocolo}
    \item \textbf{State preparation}: In a GM protocol, Alice encodes classical variables in the quadrature components $q$ and $p$, sampled from two i.i.d. Gaussian random variables $P,Q\sim\mathcal{N}(0,\tilde{V}_{Mod})$. Alice then prepares displaced coherent or squeezed states. In the case of coherent states,  $\ket{\alpha_k} = \ket{q_k + i p_k}$, where $\alpha_k = q_k + i p_k$ is a complex amplitude in phase space, with total symmetric variance of each state given by $V_q = V_p= V = 4\tilde{V}_{\text{Mod}} +1 = V_{Mod} + 1 = 2 \langle \hat{n} \rangle + 1$, where $V_{\text{Mod}} = 4\tilde{V}_{mod}$ is the quadrature operators' variance, the vacuum fluctuation is normalized to 1 in SNU, and $\langle \hat{n} \rangle = \int p_G(\alpha) |\alpha|^2d^2\alpha$. A discrete alphabet can also be used to encode classical information, rather than GM, in the case of DM protocols. 
    \item \textbf{Transmission}: The states are sent from Alice to Bob through an untrusted quantum channel, which is assumed to be fully controlled by Eve. This channel is completely characterized by two parameters: the transmittance $T$ and the excess noise $\xi$ (see Sec.~\ref{quantumchannel}).
    \item \textbf{Detection}: After receiving the channel output signals, Bob can either perform a homodyne detection, to randomly measure one of the quadratures, or a heterodyne detection, to measure both quadratures simultaneously.
    \item \textbf{Parameter estimation}: At this stage, the trusted parties already possess a correlated database of prepared and detected random variables (corresponding to asymmetric and insecure raw keys). To ensure a shared secure sequence, Alice and Bob must perform a series of classical error correction procedures using an authenticated public classical channel. The first step is parameter estimation, where Alice and Bob use part of their data to estimate the parameters characterizing the channel, i.e., the transmittance and excess noise. These parameters allow Alice and Bob to bound the information that may have leaked to Eve. If Eve's information is greater than the mutual information between Alice and Bob, the protocol is aborted.
    \item \textbf{Information reconciliation}: In this step, sophisticated error correction algorithms are applied to align the correlated data between Alice and Bob. Two approaches are possible: in direct reconciliation (DR), Alice acts as the reference and sends information derived from her data to Bob through the classical channel, which he uses to align his data with Alice's; in reverse reconciliation (RR), Bob serves as the reference instead. As discussed previously, DR is limited to a maximum transmission corresponding to $3$ dB loss; the RR process does not present a similar limitation and offers better performance \cite{Grosshans2003}.
    \item \textbf{Privacy amplification}: Finally, they perform privacy amplification to eliminate any information that Eve may have about the generated key. The result is a shorter but secure symmetric key. This process is done using universal hash functions \cite{krawczyk1994lfsr,fung2010practical,zhang2019continuous}.
\end{enumerate}
Steps 1-3 define the quantum part of the protocol, while the remaining parts represent the classical processes, known as post-processing. We remark that, in some protocols, it can be necessary to implement the sifting step at the beginning of the post-processing stage. It eliminates all uncorrelated signals between Alice and Bob. If Alice and Bob successfully perform the above steps, they can generate a secure secret key between them. 

\subsection{Prepare-and-measure and entanglement-based pictures}
\label{subsec1}

As mentioned in the first step of the previous section, the PM picture is based on the preparation of an ensemble of states. In the case of GM coherent-state protocols, Alice's density matrix is given by
\begin{equation}
\hat{\rho}_G = \dfrac{1}{\pi \langle\hat{n}\rangle} \int_{\mathbb{C}} e^{-|\alpha|^2/\langle\hat{n}\rangle}\ket{\alpha}\bra{\alpha}d^2\alpha,
\end{equation}
where $\ev{\hat{n}}=2\tilde{V}_{\text{Mod}}$. This state coincides with the thermal state, Eq.(\ref{thermal}), with total variance $V = 2\langle\hat{n}\rangle + 1$.

The security of a PM protocol can be analyzed from the perspective of an equivalent EB picture, meaning that a secure EB protocol implies a secure PM one. In this sense, Alice can work with PM or EB pictures without Bob and Eve ever noticing which one is being implemented \cite{Laudenbach2018,djordjevic2019physical}. In the EB scenario, Alice holds a purification of the state $\hat{\rho}_G$, given by a bipartite state $\ket{\phi}_{AA'}$ (in the form of Eq.(\ref{tmsvs})), from which she measures her mode ($A$) and sends the other mode ($A'$) to Bob. The associated CM before the transmission is given by
\begin{equation}
\Sigma_{AA'}^{EB} = 
\begin{pmatrix}
V \mathbb{I} & \sqrt{V^2 - 1} \sigma_z \\
\sqrt{V^2 - 1} \sigma_z & V \mathbb{I}
\end{pmatrix}.
\label{tmsvscmeq}
\end{equation}
In order to prepare an equivalent PM protocol, she must perform a heterodyne (homodyne) measurement on her mode to project Bob's mode into a coherent (squeezed) state \cite{zhang2019continuous}.

After the transmission and interaction with the untrusted quantum channel, assumed to be fully controlled by Eve, the shared state is given by
\begin{equation}
\hat{\rho}_{AB} = (\mathbb{I}_A  \otimes \mathcal{E}_{A' \rightarrow B}) (\ket{\phi}\bra{\phi}_{AA'}),
\label{finalstate}
\end{equation}
where $\mathcal{E}_{A' \rightarrow B}$ denotes the quantum channel acting on mode $A'$.

The channel is usually modeled by a bosonic phase-insensitive Gaussian channel (see Sec. \ref{quantumchannel}), well described by a beam splitter with transmissivity $T \in [0,1]$ and environment noise $\epsilon$, leading to the following transformation: $\hat{q}_B = \sqrt{T} \hat{q}_{A'} + \sqrt{1 - T} \hat{q}_E$ and $\hat{p}_B = \sqrt{T} \hat{p}_{A'} + \sqrt{1 - T} \hat{p}_E$, where $\hat{q}_E$ and $\hat{p}_E$ are the environment (Eve) quadrature operators. In terms of the statistical moments, we obtain
\begin{equation}\label{eq:CMtransformation}
\begin{split}
\boldsymbol{\bar{r}}_{AA'} &\rightarrow \mathbb{X} ~ \boldsymbol{\bar{r}}_{AA'} + \boldsymbol{r}_0, \\
\Sigma_{AA'} &\rightarrow \mathbb{X} ~ \Sigma_{AA'} ~ \mathbb{X}^T + \mathbb{Y},
\end{split}
\end{equation}
with $\mathbb{X} = \mathbb{I} \oplus  \sqrt{T}~ \mathbb{I}$, $\mathbb{Y} = 0_2 \oplus (1 - T) \epsilon ~ \mathbb{I}$, and $\boldsymbol{r}_0 =  (0_2, \sqrt{1 - T} ~ \boldsymbol{\bar{r}}_E)$ where $\mathbb{Y}$ acts only on Bob's mode, and $0_2$ is a $2 \times 2$ null matrix, and $\boldsymbol{r}_0$ is the Eve's contribution for the displacement vector, with $\boldsymbol{\bar{r}}_E$ representing Eve's mean vector. The resulting matrix has the simple form
\begin{equation}
\Sigma_{AB}^{EB} = 
\begin{pmatrix}
V \mathbb{I} & \sqrt{T(V^2 - 1)} \sigma_z \\
\sqrt{T(V^2 - 1)} \sigma_z & [TV + (1 - T)\epsilon] \mathbb{I}
\end{pmatrix}.
\end{equation}
We can further parametrize the environment noise in terms of the transmissivity and the excess noise as follows: $\epsilon = 1 + T \xi/(1 - T)$, leading to the final form of the CM after the transmission as
\begin{equation}
\Sigma_{AB}^{EB} = 
\begin{pmatrix}
V \mathbb{I} & \sqrt{T(V^2 - 1)} \sigma_z \\
\sqrt{T(V^2 - 1)} \sigma_z & [T(V-1) + 1 + T\xi] \mathbb{I}
\end{pmatrix}.
\label{eq:CMchannel}
\end{equation}
A simple derivation based on the action of a beam splitter can be found in Ref.\cite{Laudenbach2018}. It is also possible to derive Eq.(\ref{eq:CMchannel}) from the modified state $\hat{\rho}_{AB}$ by means of direct calculation of the statistical moments elements of the CM \cite{Denys2021}. In the next section, the security analysis of CV-QKD protocols is discussed, where the CM in the form of Eq.(\ref{eq:CMchannel}) is typically used as the starting point for the security proofs.

\section{Security Analysis}
\label{sec5}

Security analysis for CV-QKD protocols relies on a series of assumptions regarding the protocol implementation and the adversary’s capabilities. One of the core assumptions is that Alice and Bob strictly follow all steps of the protocol, as shown in Sec.~\ref{cvqkddescription}. 

Another fundamental assumption concerns the number of rounds Alice and Bob can perform, that is, the number of quantum states that Alice sends to Bob through the channel. In a realistic scenario, it only makes sense to assume that this number is finite. However, the experimental challenges involved make security analysis in this finite regime particularly subtle and complex \cite{Leverrier2010, Ruppert2014, Papanastasiou2018, Bauml2024, Pascual2025}. A brief introduction to this scenario is presented in Sec.~\ref{sec6}. On the other hand, introducing an idealized theoretical scenario in which the protocol is repeated an arbitrarily large number of times, even though it does not directly model a practical situation, greatly facilitates the development of tools for simplified security analysis. This scenario is commonly referred to as the asymptotic regime and will be assumed as a hypothesis throughout this section.

At first glance, this assumption may seem unnecessary or even inadequate, as in practice, Alice and Bob are subject to time, resource, and channel stability constraints and must terminate the quantum signal exchange after a certain period. Nevertheless, studies in theoretical scenarios play a crucial role by providing idealized models of the protocols, enabling the establishment of ideal upper performance bounds that serve as references for technological progress. Furthermore, these idealized models provide a foundation for assessing protocol feasibility and identifying the assumptions, constraints, and resources required for practical implementation.

The security analysis primarily focuses on attacks performed by Eve on the quantum channel, assuming she has complete control over it. In this scenario, Eve can intercept the signals, interact with them through ancillary systems, and attempt to extract information from measurements performed on these systems after the interaction. The goal of the analysis is to establish an upper bound on the amount of information Eve can gain based on the disturbance that her presence induces on the transmitted signals. This allows Alice and Bob to quantify how much of their shared information is provably secure.

In what follows, we briefly outline the key elements of CV-QKD protocols (see Sec. \ref{protocolo}) for security analysis in the asymptotic regime. Our focus is on the initial considerations that provide a more straightforward overview for readers who are beginning to explore the field of CV-QKD. In this sense, the following are fundamental aspects within the security proof: 

\begin{enumerate}
    \item \label{item:primeiro} The modulation scheme \cite{Zhang2024}, either Gaussian or discrete, and the corresponding quantum states prepared and sent by Alice. For the remainder of this section, we assume that the prepared states are coherent states\footnote{However, some of the methods in the security analysis for modulation with squeezed states are similar to the case of coherent states~\cite{zhang2019continuous}.}. This choice is motivated by the fact that coherent states are the most commonly employed in conventional CV-QKD protocols. Although the earliest proposals were formulated in terms of squeezed states, as discussed in Sec.~\ref{sec:cvqkd}, the experimental preparation of such states remains technically challenging\footnote{However, a CV-QKD implementation with squeezed light has recently been proposed \cite{tobias2025practicalcontinuousvariablequantumkey}.} \cite{Andersen_2016}. As a result, coherent state protocols have emerged as the most practical option. Nonetheless, in recent years, there has been a growing interest in exploring protocols based on squeezed and thermal states~\cite{nag2025continuous,usenko2018unidimensional, jacobsen2018complete,winnel2021minimization}.

    \item The type of detection performed by Bob on the states received from Alice after they passed through the channel \cite{morin:tel-01066655, ferraro2005, schleich2015quantum, wittmann2010, zhao2024security}. This element is essential for determining the correlation between Alice and Bob, as well as the parameters that describe the channel and the amount of information that Eve can extract. Aside from a few works that explored non-standard measurements as detection strategies for CV-QKD protocols \cite{wittmann2010, cattaneo2018, Zhao2021, zhao2024security}, most schemes employ either homodyne or heterodyne detection. Therefore, from this point forward, we will assume that Bob's detection is one of these two Gaussian measurements.
    \item The direction of information reconciliation: direct reconciliation or reverse reconciliation \cite{grosshans2002reversereconciliationprotocolsquantum}. 
    \item Finally, the communication channel (i.e., the quantum channel) considered in the protocol. As previously mentioned, it is assumed that the channel is under Eve’s control; a standard formulation of this assumption is that Eve determines the channel. However, this formulation is subtle: although Eve controls the channel, parameters that Alice and Bob can estimate, such as the channel transmissivity and excess noise, constrain the set of possible channel models. 
    
    The flexibility in modeling the channel poses one of the central challenges in security analysis: we must consider the channel model that allows Eve to extract the maximum amount of information while still being consistent with the parameter estimations obtained during the reconciliation stage.
\end{enumerate}

\subsection{Quantum channels}
\label{quantumchannel}

These four points are critical features of a CV-QKD protocol that must be considered when analyzing its security. The protocol's definition determines the first three. The fourth, however, is more delicate: although Alice and Bob may have an idea of the type of channel through which the information is transmitted, the actual structure of the channel, which corresponds to the structure of Eve’s attack, is generally unknown.

The most common assumption regarding the channel is that it is a phase-invariant Gaussian bosonic channel with transmittance 
$T$ and excess noise $\xi$. This assumption is supported by three main arguments: the first is that the optimal attack Eve can perform corresponds to a Gaussian channel~\cite{navascues2006optimality, wolf2006extremality, garcia2006unconditional} (see Theorem~\ref{Theorem 1})\footnote{This is natural in the protocols with Gaussian modulation (Sec. \ref{GaussModeSec}). As for discrete modulation, the use of this hypothesis about the channel model is based on the argument that this model describes realistic scenarios well and serves as a comparison mechanism with other protocols whenever Gaussian attacks are not proven to be optimal.}; the second is that phase-invariant channels are optimal because Alice and Bob can apply a symmetrization procedure (see Sec.~\ref{symmetrization section})\footnote{Except in some specific cases, such as unidimensional modulation~\cite{usenko2015unidimensional}}; and the third is that this hypothesis fits very well within realistic optical scenarios in the absence of Eve, as it is simply a simulation model  ~\cite{Weedbrook2012, Laudenbach2018, ghorai2019asymptotic, Denys2021explicitasymptotic, Lin2019} (e.g., optical fiber or free-space channels). The parameters $T$ and $\xi$ fully characterize the channel, and estimating them is a central aspect of the post-processing stage. For these types of channels, the state received by Bob when Alice sends a coherent state $\ket{\alpha}$, is a thermal state (see Fig.~\ref{wignerstates}) centered at $\sqrt{T}\alpha$ in phase space with variance $1+T\xi$ \cite{ghorai2019asymptotic}.

\begin{figure}[htbp]  
    \centering  
    \includegraphics[width=0.45\textwidth]{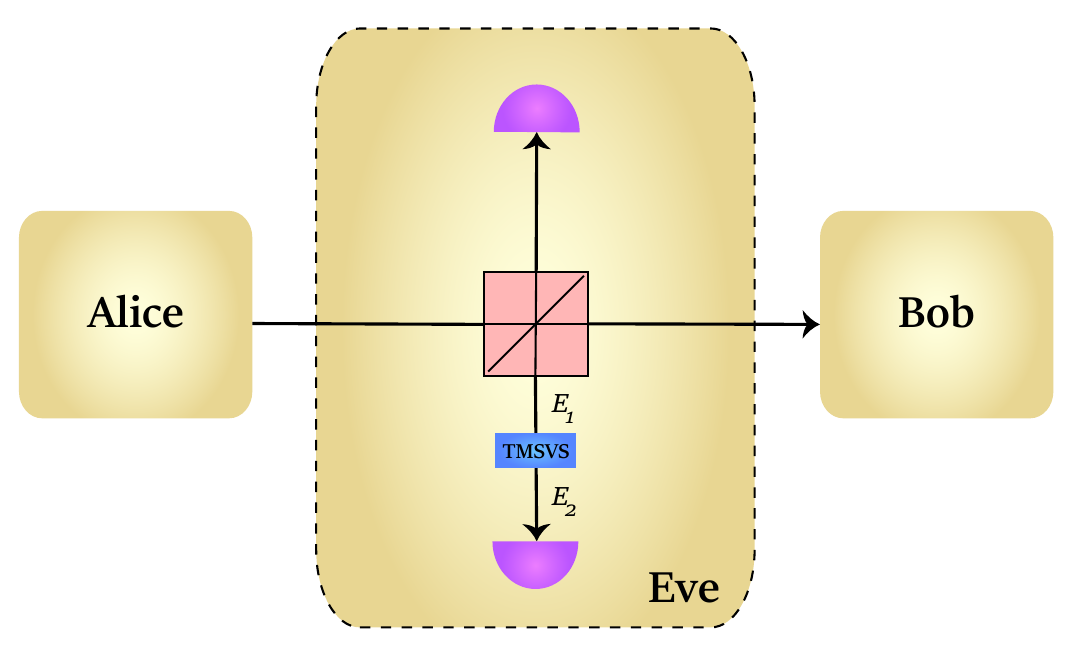}  
    \caption{Entangling cloner attack \cite{Grosshans2007}: Eve prepares a TMSVS and mixes one of its modes with the signal sent by Alice at a beam splitter. She sends one output mode of the beam splitter to Bob (which models a Gaussian bosonic channel) and keeps the other mode (ancillary system) for herself, to be measured either individually or jointly depending on the strategy she applies. }  
    \label{fig:fig0}  
\end{figure}

Since there is a general idea of how the channel should operate, and because any eavesdropper will try to remain undetected, it is possible to construct a model of how Eve might carry out her attack \cite{Grosshans2007,  Usenko2025} (see Fig.~\ref{fig:fig0}). However, there is no guarantee that the modeled attack corresponds to what actually occurs or to the most effective strategy for Eve. This is a fundamental principle in security analysis: Eve’s attack is assumed to be the one that allows her to extract the maximum amount of information without being detected.

\subsection{Eve's Strategies}
\label{subsec4}

To complete all the necessary ingredients for the security analysis, an additional assumption is needed, one that is not under the control of the trusted parties and instead depends on the adversary's capabilities \cite{Grosshans2007}. We characterize this capability as the type of attack that Eve can perform:

\begin{enumerate}
    \item  \textbf{Individual Attacks:} Eve interacts with each quantum signal separately and measures each ancillary system individually after the interaction. 
    \item \textbf{Collective Attacks:} Eve stores all ancillary systems in a quantum memory and performs an ideal collective measurement after all rounds of the protocol are completed (in particular, after post-processing) \cite{pirandola2008, Laudenbach2018}. Another critical assumption in the collective attack scenario is that Eve applies the same type of interaction in every round of the protocol, i.e., the channel remains identical for each use.

    \item \textbf{Coherent attacks:} This strategy is the most powerful class of attacks. Eve can interact simultaneously with multiple quantum signals sent from Alice to Bob. Unlike in collective attacks, she does not treat each signal individually; instead, she can perform a joint interaction on blocks of signals. In this setting, Eve can prepare and perform joint operations over all ancillary systems in a fully optimized manner. In some cases, especially when the signal states are i.i.d., coherent attacks can be reduced to collective attacks, yielding no advantage for Eve \cite{Renner05,renner2009finetti}.

\end{enumerate}

\subsection{Secure Key Rate}
\label{subsec5}

Any security proof is strictly restricted by the underlying assumptions mentioned above and holds only within that scope. Once the protocol description is specified and the hypothesis about Eve's possible attacks has been established, a theoretical security analysis can proceed.

As discussed in Sec.~\ref{cvqkddescription}, the EB version is more convenient for analyzing the security of a given protocol. In this case, Eve's interaction on the protocol is given by a purification of $\hat{\rho}_{AB}$ \cite[Cap.~5]{wilde2017b} \cite{Grosshans2007, Denys2021}:
\begin{equation}\label{eq:state}
\hat{\rho}_{ABE}=\ket{\Psi}\bra{\Psi}_{ABE} = (\mathbb{I}_A \otimes U_{A' \rightarrow BE}) \left( \ket{\Phi}\bra{\Phi}_{AA'} \right),
\end{equation}
where $U_{A' \rightarrow BE}$ is the isometric representation of the quantum channel, which introduces Eve's mode \cite{Denys2021}. 

\subsubsection{Individual attacks}

For individual attacks, Eve's attack is performed in each round of the protocol; she also measures and obtains another random variable $Z$ that is correlated with those of Alice and Bob, which, after the measurements, are given by the random variables $X$ and $Y$. Thus, the joint correlation between the three random variables can be described by the following diagonal classical-classical-classical operator \cite{Grosshans2007}
\begin{equation}\label{eq:ccc}
\hat{\rho}_{ABE}^{ccc}=\sum_{\substack{x \in \mathcal{X} \\ y \in \mathcal{Y} \\ z \in \mathcal{Z}}} p(x,y,z)\ket{x}\bra{x}\otimes \ket{y}\bra{y} \otimes \ket{z}\bra{z},
\end{equation}     
where $\mathcal{X}$, $\mathcal{Y}$, and $\mathcal{Z}$ are the alphabets of the random variables $X$, $Y$, and $Z$, respectively, i.e., the sets containing all the possible outcomes obtained from the measurements.

In this case, Eve’s information is bounded by the classical mutual information between her data and that of the
reference party (Alice or Bob, depending on the reconciliation process). Therefore the secret key rate is given by the difference between the mutual information shared by Alice and Bob and the information obtained by Eve \cite{Csiszar1978, Maurer1993}:

\begin{equation}
K_{DR} = I(X;Y) - I(X;Z),
\end{equation}
for direct reconciliation, and

\begin{equation}
    K_{RR} = I(X;Y) - I(Y;Z),
\end{equation}
for reverse reconciliation. 

\subsubsection{Collective attacks}
\label{subsecCollect}

Although individual attacks are realistic from the perspective of current technology, the fact that Eve performs her measurements in each round of the protocol introduces a significant limitation that greatly restricts her capabilities. This limitation is overcome by assuming that Eve has access to a quantum memory, where she stores the collected signals in the auxiliary system. Then, after the protocol is completed (including taking advantage of the classical information exchanged between the trusted parties), she performs collective measurements on the stored states.

This scenario is reflected in the structure of the diagonal operator in Eq.\eqref{eq:ccc}, where the ability to store the states breaks the diagonal structure of Eve's system, leading to a classical-classical-quantum state \cite{Grosshans2007, Dias2023}
\begin{equation}\label{eq:ccq}
\hat{\rho}_{ABE}^{ccq}=\sum_{\substack{x \in \mathcal{X} \\ y \in \mathcal{Y}}} p(x,y)\ket{x}\bra{x}\otimes \ket{y}\bra{y} \otimes \hat{\rho}_{E}^{xy}.
\end{equation}  
Since these attacks are more powerful than individual attacks, much of the existing literature focuses on security analysis under the assumption that Eve adopts this strategy. Henceforward, we will assume only this type of attack. 

In this case, the secret key rate is bounded by the Devetak–Winter formula \cite{Devetak2005}, given by
\begin{equation}
    K_{\text{DR}} = I(X;Y) - \chi(X;E)
\end{equation}
or
\begin{equation}\label{eq:Devetak-Winter}
    K_{\text{RR}} = I(X;Y) - \chi(Y;E),
\end{equation}
where $\chi(X;E)=S(E)-S(E|X)=S(\hat{\rho}_{E})-\int p(x)S(\hat{\rho}_{E}^x)dx$ is the Holevo information \cite{Holevo1973} between Eve's system and the part measured by Alice (or $\chi(Y;E)=S(\hat{\rho}_{E})-\int p(y)S(\hat{\rho}_{E}^y)dy$ if Bob's measurement is considered)\footnote{Note that the Holevo information is computed for the state $\rho_{XBE}=\mathcal{M}_{A\rightarrow X} (\rho_{ABE})$ where the map $\mathcal{M}_{A\rightarrow X}$ describes the measurement performed by Alice, or $\rho_{AYE}=\mathcal{M}_{B\rightarrow Y} (\rho_{ABE})$ where the map $\mathcal{M}_{B\rightarrow Y}$ describes the measurement performed by Bob \cite{Denys2021explicitasymptotic}. The integral is reduced to a summation if Alice's or Bob's variables are discrete.}. This quantity measures the amount of information that Eve can extract from the protocol. 

Since the channel is under Eve's control, the optimal attack she can perform is generally unknown. Therefore, the Devetak–Winter formula must be evaluated over all possible channels that Eve could employ. However, it is important to emphasize that the choice of possible channels in the security analysis is not arbitrary; these channels must be compatible with the parameters estimated during the parameter estimation phase. This is a subtle yet necessary adjustment to determine a secret key rate bound that does not overestimate the adversary's capabilities. Consequently, the expression of the Devetak–Winter formula in Eq.\eqref{eq:Devetak-Winter} is modified to evaluate the Holevo information over all possible attacks that Eve could perform and to consider the one that gives her the most information
\begin{equation}\label{eq:general DW}
    K_{\text{RR}} = I(X;Y) - \sup \chi(Y;E),
\end{equation} 
where $\sup$ is the supremum over all possible channels.

The first relevant aspect for proving security (\ref{item:primeiro}) plays an important role in this general definition of the key rate bound. While in the case of Gaussian modulation it is well known that the optimal attack Eve can perform is the cloning attack \cite{pirandola2008} (see Fig.~\ref{fig:fig0}), and Eq.\eqref{eq:Devetak-Winter} is sufficient to compute $K_{RR}$ (or analogously $K_{DR}$), for discrete modulation there is no known characterization of the optimal attack, and Eq.\eqref{eq:general DW} becomes necessary.

\subsection{Gaussian modulation}
\label{GaussModeSec}

In order to evaluate the secret key rate, we must compute the classical mutual information between Alice and Bob. In the case of GM, the mutual information can be written in terms of the variances as \cite{garcia2007quantum,Usenko2025}
\begin{equation}
I(X;Y) = \dfrac{\mu}{2} \log \Bigg( \dfrac{V_X V_Y}{V_X V_Y - C^2_{XY}} \Bigg),
\end{equation}
where $V_X$, $V_Y$, and $C_{XY} = \langle xy \rangle$ are the variances of Alice, Bob, and the covariance between them, respectively, with $x,y$ being the values of the variables $X$ and $Y$, and $\mu = 1(2)$ standing for homodyne (heterodyne) detection \cite{Laudenbach2018}. It can also be written as $I_{XY} = (\mu/2) \log(V_X/V_{X|Y}) = (\mu/2) \log(V_Y/V_{Y|X})$, where $V_{X|Y} = V_X -  C^2_{XY}/V_Y$ is the conditional variance.

Additionally, considering that the channel is a bosonic Gaussian channel and acts as in Eq.\eqref{eq:CMtransformation}, it follows that 
\begin{equation}
    \frac{V_X}{V_{X|Y}}=1+\mathrm{SNR}
\end{equation}   
where 
\begin{equation}
    \text{SNR}=\dfrac{T V_{\text{Mod}}}{\mu + T\xi},
\end{equation}
denotes the signal-to-noise ratio of the channel \cite{Laudenbach2018}.

Therefore, in the case of GM CV-QKD protocols, the mutual information assumes the form
\begin{equation}
I(A;B) =\dfrac{\mu}{2} \log \Bigg(1 + \dfrac{T V_{\text{Mod}}}{\mu + T\xi} \Bigg).
\label{gaussmutualinfo}
\end{equation}

In a realistic scenario, Alice and Bob cannot perfectly recover their mutual information. Therefore, the term \( I(X; Y) \) is replaced in the secret key rate by \( \beta I(X; Y) \), where \( \beta \in [0,1] \) is a parameter that quantifies the information reconciliation process efficiency and characterizes the performance of the error correction \cite{Yang2023}.

The second term of the secret key rate can be obtained considering that, as represented in Eq.\eqref{eq:state}, Eve holds a purification of the state $\hat{\rho}_{AB}$. Therefore, the entropy of $\hat{\rho}_E = \mathrm{Tr}_{AB} (\hat{\rho}_{ABE})$ is exactly the same as the entropy of $\hat{\rho}_{AB}$ \cite{araki1970entropy}.

Additionally, Bob's measurement with outcome $y$ projects the joint system of Alice and Eve onto a conditional pure state $\hat{\rho}_{AE}^y$, therefore $S(\hat{\rho}_E^y) = S(\hat{\rho}_A^y)$. Consequently, the expression for the Holevo information can be rewritten as
\begin{align} \label{eq:HolInf}
    \chi(Y;E) =& S(\hat{\rho}_{E}) - \int  p(y)S(\hat{\rho}_E^y) dy\\
    =& S(\hat{\rho}_{AB}) - \int  p(y)S(\hat{\rho}_A^y) dy.
\end{align}
For the reader’s convenience, we will denote this expression by $\chi_{\hat{\rho}_{AB}}(Y;E)$ and $K_{\text{RR}}(\hat{\rho}_{AB}) := I(X;Y) - \chi_{\hat{\rho}_{AB}}(Y;E)$.

In the protocol with Gaussian modulation, it is well known that Gaussian collective attacks are optimal \cite{garcia2006unconditional, navascues2006optimality}. This property follows from the fact that, in the asymptotic regime, coherent attacks can be reduced to collective attacks \cite{renner2009finetti}, and for collective attacks, the Gaussian extremality property is valid: 

\begin{theorem}[Gaussian extremality property \cite{wolf2006extremality,garcia2006unconditional}]\label{Theorem 1}
    For an arbitrary state $\hat{\rho}_{AB}$ with finite first and second moments,
\begin{equation}\label{eq:extremality}
K_{RR}\left(\hat{\rho}_{AB}^G\right) \leq K_{RR}\left(\hat{\rho}_{AB}\right),
\end{equation} 
where $\hat{\rho}^G_{AB}$ denotes the Gaussian state with the same covariance matrix as $\hat{\rho}_{AB}$.
\end{theorem}
This is a powerful and frequently used tool in security analysis.  

Since a Gaussian state is fully characterized by its covariance matrix, its entropy can be computed directly from the symplectic eigenvalues of that matrix \cite{navascues2005, Laudenbach2018}. Thus, we can write
\begin{equation}\label{eq:Holevo}
    \chi_{\hat{\rho}_{AB}^G}(Y;E) = g\left(\nu_1 \right) + g\left(\nu_2 \right) - g\left(\nu_3\right),
\end{equation}
where \( \nu_1, \nu_2 \) are the symplectic eigenvalues of the covariance matrix \( \Sigma_{AB}^{EB} \), which, in the case of Gaussian modulation, is given by Eq.\eqref{eq:CMchannel}. In particular, $\nu_1,\nu_2$ are given by Eq.\eqref{eq:gaussian ev}.

On the other hand, the symplectic eigenvalue $\nu_3$ depends entirely on the type of measurement performed by Bob, which is naturally embedded in the conditional entropy  $S(A|Y) = \int p(y)\, S(\hat{\rho}_A^y)\, dy$.

In the case of conventional measurements, the calculation of $\int p(y)\, S(\hat{\rho}_A^y)\, dy$ can be replaced by the evaluation of \(g(\nu_3)\), where \(\nu_3\) is the symplectic eigenvalue of the average conditional covariance matrix of Alice's subsystem after Bob's measurement, \(\Sigma_{A|Y}\), which is given by\cite{Laudenbach2018, Weedbrook2012,Zhang2024}
\begin{equation}
    \Sigma_{A|Y}=\gamma_{A}-\gamma_{AB}\left(\Pi_{\hat{q},\hat{p}}\gamma_B\Pi_{\hat{q},\hat{p}}\right)^{-1}\gamma_{AB}^{T},
\end{equation}
for homodyne detection, where $\Pi_{\hat{q}}=\text{diag}(1,0)$ and $\Pi_{\hat{p}}=\text{diag}(0,1)$. For heterodyne detection, the CM is given by
\begin{equation}
    \Sigma_{A|Y}=\gamma_{A}-\gamma_{AB}\left(\gamma_B+\mathbb{I}\right)^{-1}\gamma_{AB}^{T}.
\end{equation}

In terms of the elements of the standard form of the CM (Eq.\eqref{standardform}), $\nu_3$  is given by \cite{Laudenbach2018,Denys2021}
\begin{equation}
     \nu_3=\sqrt{a\left(a-\frac{c^2}{b}\right)},
\end{equation}
for homodyne detection, or

 \begin{equation}
    \nu_3 = a - \frac{c^2}{b+1},
 \end{equation}
for heterodyne detection. For direct reconciliation, the expressions are analogous. We do not delve into direct reconciliation, as it is not a widely used strategy compared to reverse reconciliation. This is due to the additional advantage that reverse reconciliation offers by breaking the symmetry between Bob and Eve, which allows for gains in terms of achievable distance. In contrast, with direct reconciliation, the symmetry between Bob and Eve results in Eve having more information than Bob when the channel has losses greater than 50\% \cite{Grosshans2007}. 

\subsection{Trusted noise model}
\label{subsec6}

As discussed so far, the excess noise $\xi$ is a general noisy parameter originating from all possible sources. The most paranoid approach to QKD assumes that all information losses and noises are due to Eve's presence. The security analysis based on this assumption guarantees worst-case performance for Alice and Bob, given the type of Eve's attack (see Sec.~\ref{subsec5}). However, while quantum mechanics fundamentally limits Eve's power, her effective power also depends on one's assumption about her technological abilities. In this sense, this completely untrusted noise assumption may overestimate her potential information and negatively impact the protocol's performance. Since there is no device that is free from imperfections in real experimental implementation, in a trusted device scenario, it may be reasonable to assume that Eve has no access to all noise sources \cite{lodewyck2007quantum,fossier2009improvement,garcia2009continuous,usenko2016trusted,laudenbach2019analysis}.

In the so-called trusted noise model of QKD, one may assume, for instance, that Bob's lab is isolated from Eve \cite{usenko2016trusted}. Within this assumption, the detector's noise and efficiency may be excluded from the total excess noise attributed to Eve, which, in turn, leads to higher secret key rates. Let's first rewrite Eq.(\ref{eq:CMchannel}) explicitly in terms of the detector's efficiency, omitting from hereafter the EB superscript for the rest of this work
\begin{equation}
\Sigma_{AB} = 
\begin{pmatrix}
V \mathbb{I} & \sqrt{\eta ~ T_{ch}(V^2 - 1)} \sigma_z \\
\sqrt{\eta ~ T_{ch} (V^2 - 1)} \sigma_z & [\eta ~ T_{ch} (V-1) + 1 + \eta ~ T_{ch} \xi] \mathbb{I}
\end{pmatrix},
\end{equation}
where we use that $T = \eta T_{ch}$, with $\eta$ being the detector's efficiency, and $T_{ch}$ is the channel transmissivity, which, for a typical fiber, can be written as $T_{ch} = 10^{-\gamma d/10}$, with $\gamma = 0.2$ dB/km and $d$ being the transmission distance. Remember that, in this completely untrusted noise model, $\xi$ comprises all noise sources. Considering the trusted noise model, Eve is isolated from Bob's lab; thus, the initial CM does not involve both detectors' efficiency and their noise contribution \cite{djordjevic2019physical,laudenbach2019analysis}
\begin{equation}
\Sigma_{AB} = 
\begin{pmatrix}
V \mathbb{I} & \sqrt{T_{ch}(V^2 - 1)} \sigma_z \\
\sqrt{T_{ch} (V^2 - 1)} \sigma_z & [T_{ch} (V-1) + 1 + T_{ch} \xi_{ch}] \mathbb{I}
\end{pmatrix},
\label{cmtrusted}
\end{equation}
where $\xi_{ch} = \xi_{mod} + \xi_{Ram} + \xi_{phase} + \xi_{PR} + \cdots$, for the modulation, Raman, phase, and phase-recovery noises, among other sources, excluding Bob's detection electronic noise, $\xi_{\text{el}}$, which can reach $60\%$ of the total noise \cite{Laudenbach2018}. The mutual information, Eq.(\ref{gaussmutualinfo}), is also rewritten considering only $T_{ch}$ and $\xi_{ch}$.

\begin{figure}[t!]
    \centering
    \includegraphics[width=1\linewidth]{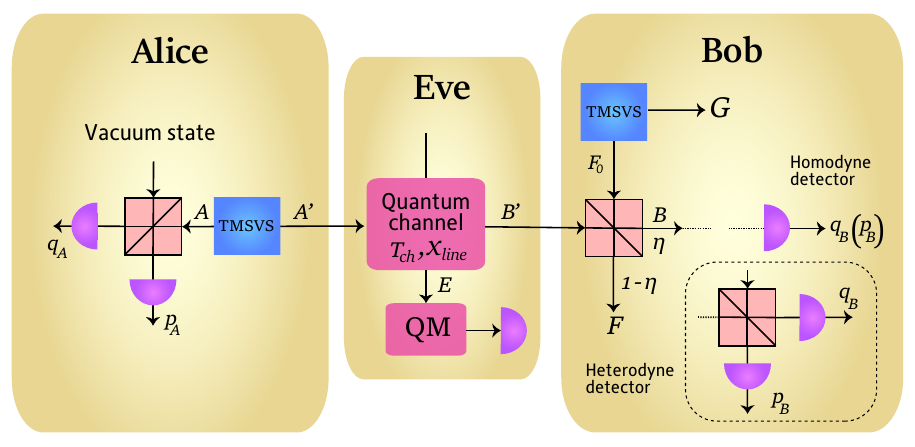}
    \caption{Trusted noise model for a GM coherent state CV-QKD protocol. Alice and Bob share a TMSVS, in which Alice performs a heterodyne detection on her mode $A$ and sends the other mode to Bob through the quantum channel. Eve interacts with this mode, leading to an output mode $B'$, keeping her results in a quantum memory (QM). This mode interacts with one mode from another TMSVS by mixing them in a beam splitter. The resulting mode $B$ is either measured by a homodyne or heterodyne detection.}
    \label{trustednoise}
\end{figure}

In order to model the detector's efficiency and noise independently of other sources, an auxiliary TMSVS system and a beam splitter are required at Bob's lab. This extra system is necessary to model the electronic noise originating from the detector\footnote{By adding this extra TMSVS, we can model thermal noise by tracing out one mode and ensure that the overall state remains pure.}, while the beam splitter represents the detector's efficiency. In Fig.~\ref{trustednoise}, we show a schematic representation of a trusted noise model for a coherent state-based GM CV-QKD protocol. In this protocol, as usual, Alice keeps her mode $A$ from the shared TMSVS to perform her heterodyne measurement and sends the mode $A'$ through the quantum channel with transmissivity $T_{ch}$ and total channel noise $\chi_{\text{line}} = 1/T_{ch} - 1 +\xi_{ch}$, accounting for the attenuation effect and the excess noise $\xi_{ch}$, attributed to Eve. Bob's input mode, denoted as $B'$, is mixed in the beam splitter with the mode $F_0$ from the auxiliary TMSVS at Bob's lab, resulting in the output mode $B$, in which he performs a homodyne or heterodyne measurement. The noise contribution from Bob's apparatus can be written as
\begin{equation}
\chi_{\text{det}} = 
\begin{cases}
\chi_{\text{hom}} = (1 - \eta + \xi_{\text{el}})/\eta \\
\chi_{\text{het}} = (2 - \eta + 2\xi_{\text{el}})/\eta,
\end{cases}
\end{equation}
leading to a total noise of
\begin{equation}
\chi = \chi_{\text{line}} + \dfrac{\chi_{\text{det}}}{T_{ch}}.
\end{equation}

Due to the extra modes taken into account in this scenario, there will be some extra symplectic eigenvalues to be considered in the von Neumann entropy, Eq.(\ref{eq:g function}), when evaluating Eve's information.

Using purification arguments, we can still compute the first term of the Holevo information as shown in Eq.(\ref{eq:HolInf}), from the state in Eq.(\ref{cmtrusted}) written as
\begin{equation}
\Sigma_{AB} = 
\begin{pmatrix}
V \mathbb{I} & \sqrt{T_{ch}(V^2 - 1)} \sigma_z \\
\sqrt{T_{ch} (V^2 - 1)} \sigma_z & T_{ch} (V + \chi_{\text{line}})) \mathbb{I}
\end{pmatrix}.
\end{equation}

The second term is computed from the state $\hat{\rho}_{AFG}^y$, which describes the total state after Bob's projective measurement, which is obtained from partial measurements applied to the composed CM
\begin{equation}
\Sigma_{AFGB} =
\begin{pmatrix}
\gamma_{AFG} & \gamma_{AFGB} \\
\gamma_{AFGB}^T & \gamma_B
\end{pmatrix}.
\label{cmtrusted2}
\end{equation}
This state is derived by rearranging the rows and columns of the state
\begin{equation}
\Sigma_{ABFG} = \mathcal{S}_{\hat{BS}} ~ (\Sigma_{AB} \oplus \Sigma_{F_0G}) ~ \mathcal{S}^T_{\hat{BS}},
\end{equation}
where $\Sigma_{F_0G}$ describes the auxiliary TMSVS at Bob's lab used to model the detector's noise with variance $\omega$, representing the thermal noise, given by
\begin{equation}
\Sigma_{F_0G} =
\begin{pmatrix}
\omega~\mathbb{I} & \sqrt{\omega^2 -1} ~ \sigma_z \\
\sqrt{\omega^2 -1} ~ \sigma_z & \omega~\mathbb{I}
\end{pmatrix},
\end{equation}
and, in this case, the detector's imperfection is modeled by the beam splitter operator as described by $\mathcal{S}_{\hat{BS}} = \mathbb{I}_A \otimes \mathcal{S}_{\hat{BS}_{B'F_0}} \otimes \mathbb{I}_B$, with
\begin{equation}
\mathcal{S}_{\hat{BS}_{B'F_0}} = 
\begin{pmatrix}
\sqrt{\eta} ~ \mathbb{I} & \sqrt{1 - \eta} ~ \mathbb{I}\\
-\sqrt{1 - \eta} ~ \mathbb{I} & \sqrt{\eta} ~ \mathbb{I}.
\end{pmatrix}.
\end{equation}
A detailed derivation of the state in Eq.(\ref{cmtrusted2}) can be found in chapter 8 of Ref.\cite{djordjevic2019physical}.

Finally, applying partial measurement to the state in Eq.(\ref{cmtrusted2}), we obtain
\begin{equation}
    \Sigma_{AFG|Y}=\gamma_{AFG}-\gamma_{AFGB}\left(\Pi_{\hat{q},\hat{p}}\gamma_B\Pi_{\hat{q},\hat{p}}\right)^{-1}\gamma_{AFGB}^{T},
\end{equation}
for homodyne detection and
\begin{equation}
    \Sigma_{AFG|Y}=\gamma_{AFG}-\gamma_{AFGB}\left(\gamma_B+\mathbb{I}\right)^{-1}\gamma_{AFGB}^{T},
\end{equation}
for the heterodyne case.

\begin{figure}[t!]
    \centering
    \includegraphics[width=1\linewidth]{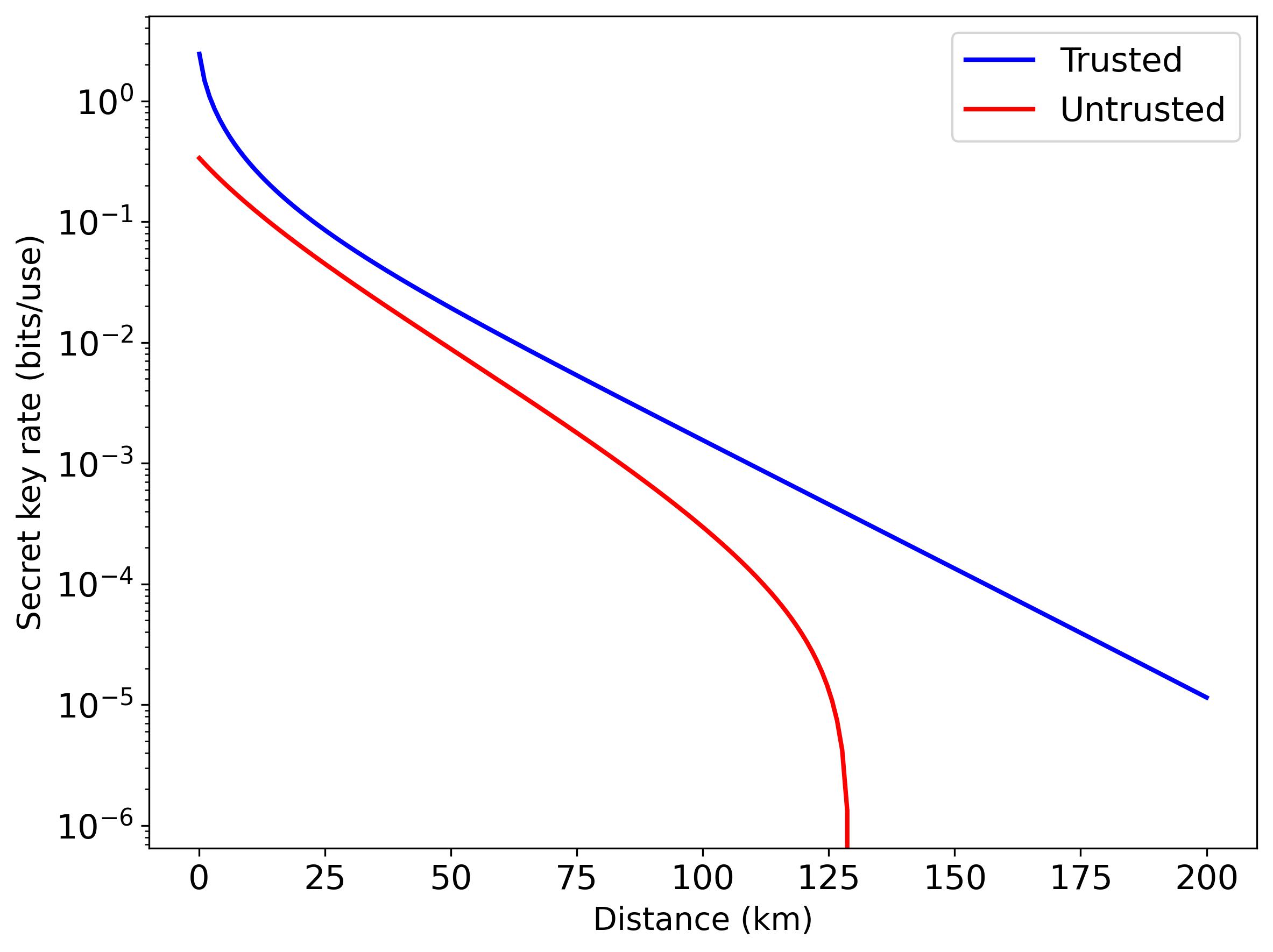}
    \caption{Secret key rate for the No-switching protocol in both untrusted and trusted models. We used viable values for the parameters: $\beta = 0.95$ $\xi = 0.05$, $\xi_{ch} = 0.02$, $\xi_{el} = 0.03$, and $\eta = 0.6$.}
    \label{trustuedskrplot}
\end{figure}

The general form of the symplectic eigenvalues of $\Sigma_{AFG|Y}$ can be written as
\begin{equation}
\nu_{3,4} = \sqrt{\frac{1}{2} ( C \pm \sqrt{C^2 - 4 D})}, ~~~~ \nu_5 = 1,
\label{eq:ev2}
\end{equation}
where
\begin{equation}
\begin{split}
C_{\mathrm{hom}}
 &= \frac{\chi_{\mathrm{hom}}\Delta + T_{ch}\!\left(V+\chi_{\text{line}}\right) + V\sqrt{\Gamma}}
        {T_{ch}\!\left(V+\chi\right)} , \\
D_{\mathrm{hom}}
 &= \sqrt{\Gamma}\;
   \frac{\sqrt{\Gamma}\,\chi_{\mathrm{hom}} + V}
        {T_{ch}\!\left(V+\chi\right)}, \\
C_{\mathrm{het}}
 &= \frac{1}{T_{ch}^{2}\!\left(V+\chi\right)^{2}} \Bigg(2\chi_{\mathrm{het}}\!\left[ V\sqrt{\Gamma} + T_{ch}\!\left(V+\chi_{\text{line}}\right) \right] \\
   &+ \Delta\,\chi_{\mathrm{het}}^{2} + \Gamma + 1 + 2T_{ch}(V^2 - 1)\Bigg)\\
D_{\mathrm{het}}
 &= \left(
     \frac{V + \sqrt{\Gamma}\,\chi_{\mathrm{het}}}
          {T_{ch}\!\left(V+\chi\right)}
   \right)^{2},
\end{split}
\end{equation}
for the homodyne and heterodyne detections, respectively. In the equations above, $\Delta$ and $\Gamma$ are the parameters specified in Eq.(\ref{eq:ev}). From these results, the Holevo information in the trusted noise model can be computed as $\chi_{\hat{\rho}_{AB}^G}^{\text{trusted}}(Y;E) = g\left(\nu_1 \right) + g\left(\nu_2 \right) - g\left(\nu_3\right) - g\left(\nu_4\right)$, since $g(\nu_5) = 0$. In Fig.~\ref{trustuedskrplot}, we show the gain of the trusted over the untrusted model for the GM coherent state-based protocol (No-switching) for viable parameters: $\beta = 0.95$ $\xi = 0.05$, $\xi_{ch} = 0.02$, $\xi_{el} = 0.03$, and $\eta = 0.6$. It is clear that the advantage of the trusted noise model in terms of maximum transmission distance and higher secret key rate over short distances.

While our example and most of the literature apply this model in GM CV-QKD protocols, it is also applicable to DM protocols \cite{almeida2021secret,almeida2023reconciliation,becir2012continuous}.

\subsection{Discrete modulation}
\label{subsec:DM}

Although the Gaussian modulation protocol is optimal and its security is proven, it faces practical limitations. Until 2010 \cite{leverrier2010continuous}, with the reconciliation schemes available, these protocols suffered from a significant drop in efficiency in low SNR regimes, which are precisely the regimes required for distributing secret keys over long distances. This problem was addressed by introducing protocols with discrete modulation, where Alice prepares states from a finite set according to a discrete probability distribution.  

Nowadays, there are already several works demonstrating high-efficiency reconciliation for Gaussian modulation protocols \cite{pacher2016information,Wang2018HighSpeed,zhang2020,Yang2023,leverrier2023informationreconciliationdiscretelymodulatedcontinuousvariable}. Nevertheless, one advantage still preserved by discrete-modulation protocols is their greater compatibility with classical communication systems \cite{Eriksson2019WDM,LopezGrande:21}. However, there is no closed formula for DM mutual information. The usual approach involves calculating the secret key rate using the Gaussian mutual information, Eq.\eqref{gaussmutualinfo}, which is a good approximation for low values of SNR \cite{Wu2010,Denys2021}. An alternative approach, which remains valid even for high SNR values, is to compute it directly from the definition of mutual information, Eq.(\ref{mutualinfo}), through numerical evaluation. An example is given in the following, Eq.(\ref{mutualinfodisc}).

\subsubsection{Pure-loss channel}
In some cases, when Eve’s optimal attack is known, the Holevo information can be directly computed from the state that models this attack. For example, if we consider the pure-loss channel, Eve's information can be directly obtained from the vacuum mode that enters in a beam splitter with transmittance \( T \) \cite{Grosshans2005pureloss}. In this case, if Alice sends the coherent state $\ket{\alpha_k}$, Eve obtains the coherent state $\ket{\sqrt{1-T}\,\alpha_k}$ and Bob receives the coherent state $\ket{\sqrt{T}\, \alpha_k}$. Therefore, for reverse reconciliation, we have 
\begin{equation}\label{holevonoGE}
    \chi(Y;E) = S(\hat{\rho}_E) - \int p(y)S(\hat{\rho}_E^y)dy,
\end{equation}
where
\begin{equation}
    \hat{\rho}_E = \sum_{k} p_k \, \ket{\sqrt{1-T}\,\alpha_k}\bra{\sqrt{1-T}\,\alpha_k}
\end{equation}
and 
\begin{equation}
    \hat{\rho}_{E}^y=\sum_{k}p(\alpha_k|y)\ket{\sqrt{1-T}\,\alpha_k}\bra{\sqrt{1-T}\,\alpha_k},
\end{equation}
with \(p_k = p(\alpha_k)\) being the probability that Alice prepared the state \(\ket{\alpha_k}\), and 
\begin{equation}
    p(\alpha_k|y) = \frac{p(\alpha_k) \, p(y|\alpha_k)}{p(y)},
\end{equation}
where \(p(y)\) is the probability that Bob measures the outcome \(y\) and it depends on the measurement that Bob chooses to perform \cite{Sych2010,Heid2006,notarnicola2023hybrid,Dias2023}. 

This framework facilitates the direct computation of the Holevo quantity for both  reverse and direct reconciliation (in the case of direct reconciliation, we have \cite{Heid2006}
\(
S(\hat{\rho}_E^x) = 0,
\)
so that
\(
\chi(X;E) = S(\hat{\rho}_E)
\)), thus allowing for a straightforward application of the Devetak-Winter formula, Eq.\eqref{eq:Devetak-Winter}. 

Finally, the key rate calculation can be completed through the evaluation of the mutual information, Eq.\eqref{mutualinfo}, between Alice’s string 
$X$ and Bob’s measurement outcome string $Y$, which can be obtained from the following equation\footnote{The conditional entropy $H(X|Y)$ is expressed as an integral over $Y$, reflecting the continuous nature of Bob's measurement outcomes.}: 
\begin{equation}
\begin{split}
I(X; Y) &= H(X) - H(X|Y)\\
&= -\sum_{k=1}^{n} p_k \log(p_k) - \int p(y)\, H(X|Y=y)\, dy.
\label{mutualinfodisc}
\end{split}
\end{equation}

Early security analyses usually embraced the simplifying assumption of purely lossy channels. One of the pioneering works under this hypothesis was carried out by Heid and Lütkenhaus \cite{Heid2006}, who analyzed a protocol with binary modulation in which Bob performs heterodyne measurements. Although initially developed for binary constellations, the model was later generalized to larger constellations \cite{Sych2010, Kanitschar2022}.

\subsubsection{Arbitrary channels for BPSK and 3-PSK}
One of the first approaches to consider the presence of excess noise in the quantum channel was presented by Zhao \textit{et al.} \cite{zhao2009asymptotic} for binary modulation (BPSK) and homodyne detection under collective attacks. In this work, the Devetak-Winter formula is rewritten using entropic properties, and in particular, the Holevo information term can be expressed as
\begin{align}
    \chi(Y;E) =& S(E) - S(E|Y) \nonumber \\
     =& S(E|X) + \chi(X;E) - S(E|Y),
\end{align}
after which bounds are calculated for each expression using two properties of the binary entropy: it is a decreasing and concave function related to the overlap of non-orthogonal quantum states. The generalization of this method to constellations with three coherent states was developed by Brádler and Weedbrook \cite{bradler2018security}, involving a considerably more intricate analysis regarding the validity of the two entropy properties required to apply the method originally proposed for two states. 

This analysis is too complicated to be applied to modulations with more states. Therefore, more sophisticated or simplified strategies are of great importance for the development of security proofs.   

\subsubsection{Gaussian extremality property}

Unlike Gaussian modulation protocols, for the discrete modulation case, the optimal attacks that Eve can perform are generally unknown. As a result, the secret key rate is bounded by the general formula, Eq.\eqref{eq:general DW}, since the Holevo information must be estimated as the maximum possible value, without knowing which channel model would allow Eve to obtain the most information.

Since we are working in an infinite-dimensional Hilbert space, computing \( \sup \chi(Y;E) \) is not trivial, and methods that can overcome the dimensionality problem are highly valued. In this context, the Gaussian extremality property plays a crucial role, and many of the results in the security analysis of discrete modulation protocols rely on strategies that make use of this result. In the EB protocol with Gaussian modulation corresponding to Eve's optimal attack, the state shared by Alice and Bob $\hat{\rho}_{AB}$ is the Gaussian state obtained from the TMSVS in Eq.\eqref{tmsvs} after the application of a thermal loss channel on the second mode.

In the case of discrete modulation, $\hat{\rho}_{AB}$ is generally unknown; thus, its Gaussianity is not guaranteed. Nevertheless, an immediate consequence of Theorem \ref{Theorem 1}  provides an upper bound on the Holevo information: 

\begin{corollary}[Gaussian extremality property \cite{wolf2006extremality,LeverrierPhD2009}]\label{Corolario 1}
    For an arbitrary state $\hat{\rho}_{AB}$ with finite first and second moments,
\begin{equation}\label{eq:extremality}
\chi_{\hat{\rho}_{AB}^G}(Y;E) \geq \chi_{\hat{\rho}_{AB}}(Y;E),
\end{equation}
where $\hat{\rho}^G_{AB}$ denotes the Gaussian state with the same covariance matrix as $\hat{\rho}_{AB}$.
\end{corollary}
However, the quality of this bound clearly depends on how Gaussian the shared state is.  

The application of the Gaussian extremality property is not straightforward for a general channel, as it requires an additional reformulation of the problem. 

\subsubsection{Symmetrization Approach}
\label{symmetrization section}

For discrete modulation protocols, the ensemble is formed by a finite set of coherent states (although the protocols are not restricted to this type of states), whose statistical mixture defines the average state of the constellation
\begin{equation} \label{eq:tau_state}
    \tau = \sum_k p_k \ket{\alpha_k}\bra{\alpha_k},
\end{equation}
where the states $\ket{\alpha_k}$ are coherent states prepared by Alice with probability $p_k$. A purification of this state is given by
\begin{equation}\label{eq:purification}
    \ket{\Phi}_{AA'} = \sum_k \sqrt{p_k} \ket{\psi_k} \ket{\alpha_k},
\end{equation}
where $|\psi_k\rangle := \sqrt{p_k}\, \bar{\tau}^{-1/2} |\bar{\alpha}_k\rangle$ (see \cite{Denys2021}\footnote{Here $\bar{\tau} = \sum_k p_k |\bar{\alpha}_k\rangle\langle\bar{\alpha}_k|$ is the density matrix of the phase-conjugated coherent states, with $\bar{\alpha}_k$ being the complex conjugate of $\alpha_k$.}). Eq.\eqref{eq:purification} indicates that Alice performs a projective measurement on her part of the system in the basis of the states $\ket{\psi_k}$. The state shared by Alice and Bob, $\hat{\rho}_{AB} = (\mathbb{I}_A \otimes \mathcal{E}_{A' \rightarrow B})(\ket{\Phi}\bra{\Phi}_{AA'})$, is determined by the choice of the channel $\mathcal{E}_{A'\rightarrow B}$. Then, the supremum from Eq.\eqref{eq:general DW} is taken over all possible states $\hat{\rho}_{AB}$ 
\begin{equation}\label{eq:DW}
    K=\beta I(X;Y)-\sup_{\hat{\rho}_{AB}}\chi_{\hat{\rho}_{AB}}(Y;E).
\end{equation}

After applying the protocol \(n\) times (under the assumption of collective attacks), Alice and Bob share \(N\) copies of the state, i.e., \(\hat{\rho}_{AB}^{\otimes N}\)\cite{Grosshans2007, Weedbrook2012}. The possibility of applying the Gaussian states extremality argument depends on reformulating the problem of determining the supremum of the Holevo information in Eq.\eqref{eq:DW}.

The application of the Gaussian extremality argument relies only on the knowledge of the second statistical moment of \(\hat{\rho}_{AB}\). By applying the symmetrization map

\begin{equation}
    \text{Sym}(\hat{\rho}_{AB}^{\otimes N}) = \int_{U \in \mathcal{G}} (U \otimes U^*) \, \hat{\rho}_{AB}^{\otimes N} \, (U \otimes U^*)^{\dagger} \, dU
\end{equation}
to the state \(\hat{\rho}_{AB}^{\otimes N}\), where $\mathcal{G}$ is the group of passive transformations in the phase space, and $dU$ is the Haar measure over $\mathcal{G}$, the resulting state $\text{Sym}(\hat{\rho}_{AB}^{\otimes N})$ is phase-invariant with respect to joint $U\otimes U^*$ transformations. Moreover, it eliminates all asymmetries that the state \(\hat{\rho}_{AB}\) might have in phase space, and its second-order statistics become symmetric (identical to a Gaussian state)\cite{Leverrier2012, LeverrierPhD2009}. This implies that the covariance matrix of \(\hat{\rho}_{AB}\) (in the Eq.\eqref{eq:CMtransformation}) assumes a symmetric form that depends only on three parameters \cite{leverrier2015composable,Denys2021,LeverrierPhD2009}
\begin{equation}\label{eq: sym CM}
\Sigma_{sym} = \begin{pmatrix}
    V\mathbb{I} & Z\sigma_z\\
    Z\sigma_z & W\mathbb{I}
\end{pmatrix},
\end{equation}
with 
\begin{align}
V &:= \frac{1}{2}\left(\langle \hat{x}_A^2 \rangle_{\hat{\rho}} + \langle \hat{p}_A^2 \rangle_{\hat{\rho}}\right), \\
W &:= \frac{1}{2}\left(\langle \hat{x}_B^2 \rangle_{\hat{\rho}} + \langle \hat{p}_B^2 \rangle_{\hat{\rho}}\right), \\
Z &:= \frac{1}{4} \left( \langle \{\hat{x}_A, \hat{x}_B\} \rangle_{\hat{\rho}} - \langle \{\hat{p}_A, \hat{p}_B\} \rangle_{\hat{\rho}} \right),
\end{align}

A justification for applying the $\text{Sym}$ map is that, in the PM protocol, Alice and Bob can apply a random orthogonal transformation to the classical data $X$, and $Y$, and then forget which one was performed~\cite{Leverrier2012,LeverrierPhD2009}, without compromising security.

Reconciliation is optimized for a Gaussian channel\footnote{In CV-QKD protocols, the channel most commonly encountered in experimental implementations (particularly in optical fiber transmissions) is typically modeled as an additive white Gaussian noise (AWGN) channel \cite{Denys2021explicitasymptotic}.}, meaning that the random variable \( Y \) is modeled as \cite{LeverrierPhD2009}
\begin{equation}
Y = t X + \Xi,    
\label{eq:channel_model}
\end{equation}
where \( t \) is a transmission factor, and \( \Xi \) is a random variable modeling the added noise, characterized by its variance \( \sigma^2 \). Therefore, the reconciliation procedure is unaffected if Alice and Bob both apply the same random orthogonal transformation \( R \) to their respective data, since
\begin{equation}
R Y = t R X + \Xi',
\end{equation}
where \( \Xi' \) is a rotated noise with the same variance \( \sigma^2 \). 

This symmetrization process mixes the quadratures such that the information originally separated in \(\hat{q}\) or \(\hat{p}\) becomes distributed across combinations of both. Consequently, the symmetrization makes the protocol in which Bob performs homodyne detection by randomly choosing between the \(\hat{q}\) and \(\hat{p}\) quadratures equivalent\footnote{No information is discarded in the case of homodyne detection.} to the protocol with heterodyne detection.

An important point is that even though the state \( \text{Sym}(\hat{\rho}_{AB}^{\otimes N}) \) exhibits characteristics similar to those of a Gaussian state, it is not necessarily close to one. However, the symmetrization ensures that the resulting state can be better approximated by a symmetric Gaussian state, which enables the application of the Gaussian extremality property. Nevertheless, in some scenarios, typically involving a small number of states, the Gaussian extremality assumption does not yield tight bounds (see Fig.~\ref{fig:grafica}).

\subsubsection{Semidefinite programming}
In the covariance matrix of Eq.\eqref{eq: sym CM}, the parameter \(V\) does not depend on the channel, only on the modulation, while \(W\) can be obtained from Bob's measurements. The parameter \(Z\), on the other hand, determines the correlations between Alice’s and Bob’s modes, which depend on the channel and are therefore beyond the control of both Alice and Bob\footnote{In the PM protocol, the state \(\hat{\rho}_{AB}\) (shared by Alice and Bob in the EB protocol) is just a mathematical object that is conveniently used in the security analysis but is unknown in the actual implementation.}.

Additionally, the Holevo information for the Gaussian state with covariance matrix \(\Sigma_{sym}\), computed from Eq.\eqref{eq:Holevo}, is a decreasing function of the parameter \(Z\)  \cite{leverrier2015composable,Denys2021}. This is a crucial property when applying the Gaussian extremality property, since determining the supremum of the Holevo information over all possible channels can be translated into finding the minimal possible value of the parameter $Z$, which depends on the possible state (described by a density operator\footnote{$\hat{\rho}\succeq 0$ i.e., is positive semidefinite} $\hat{\rho}$) shared by Alice and Bob.

In this regard, the work of Ghorai \textit{et al.} \cite{ghorai2019asymptotic} addressed the security analysis of a four-coherent-state (QPSK\footnote{In each round of the protocol , Alice prepares and sends to Bob one of the four coherent states: $\ket{\alpha_k}=\ket{e^{k i \pi/2}\alpha}$, $k=0,1,2,3.$} \cite{leverrier2009}) protocol using heterodyne measure and assuming coherent attack, providing a very clear example of how to use the Gaussian extremality property in security proofs. A key point in their proof is the procedure by which the set of possible channels (restricted to those consistent with the parameters estimated during the post-processing phase) is translated into conditions for minimizing the parameter $Z$. Since $Z$ is a functional of a positive semidefinite operator, the optimization of the correlation parameter can be carried out via a semidefinite program (SDP). The convex optimization problem for the QPSK protocol was defined as follows:
\begin{align}\label{SDP}
\nonumber\text{minimize:} \quad & Z(\hat{\rho})\\ 
\text{subject to:} \quad
&\begin{cases}
 &\mathrm{Tr}_{B}(\hat{\rho}) = \mathrm{Tr}_{A'} \left( |\Phi\rangle\langle\Phi|_{AA'} \right),\\
& \mathrm{Tr}(B_0\hat{\rho}) = v,\\  
& \mathrm{Tr}(B_1\hat{\rho}) = c, \\
& \hat{\rho} \succeq 0,
\end{cases}
\end{align}
where $Z(\hat{\rho})=\frac{1}{4} \left( \langle \{\hat{q}_A, \hat{q}_B\} \rangle_{\hat{\rho}} - \langle \{\hat{p}_A, \hat{p}_B\} \rangle_{\hat{\rho}} \right)$. The parameter
$v$ quantifies the variance of Bob’s measurement outcomes, while $c$
quantifies the correlation between the states prepared by Alice and Bob’s measurements, where \(B_0\) and \(B_1\) denote the operators associated with the estimation of these two quantities \cite{ghorai2019asymptotic}. For phase-invariant Gaussian channel, $v=1 + 2T \alpha^2 + T \xi$ and $c=2\sqrt{T}\alpha$ (see Observation \ref{observation1}).  

\begin{Observation}\label{observation1}
    In the PM protocol, Bob receives from Alice the state $\mathcal{E}(\ket{\alpha_k}\bra{\alpha_k})$. This procedure can also be explained in another way: Alice chooses one label corresponding to each state she sends and encodes her choice into a classical variable. For example, since the states are sent along the axes in phase space, it is possible to choose the labels $+1$ or $-1$ when Alice sends $\ket{\alpha}$ or $\ket{-\alpha}$, respectively, and analogously to choose $+1$ or $-1$ when Alice sends $\ket{i\alpha}$ or $\ket{-i\alpha}$, respectively. Then, we can compute the parameter $c$ as the covariance between this classical choice ($x_0=1$ or $x_1=-1$) and the corresponding quadrature measurements performed by Bob on the received quantum state, which, as mentioned earlier, is a thermal state centered at $\sqrt{T}\alpha_k$ with variance $1+T\xi$. Therefore, 
    \begin{align}
        c&=\frac{1}{4}\left[ \sum_{i=0}^1 x_{i}\langle \hat{q} \rangle_{\mathcal{E}( \ket{\alpha_{2i}}\bra{\alpha_{2i}})}+\sum_{i=0}^1 x_i\langle \hat{p} \rangle_{\mathcal{E}( \ket{\alpha_{2i+1}}\bra{\alpha_{2i+1}})}\right]\\ &=\frac{1}{4}\left[2Re\left(\sqrt{T}\alpha\right)-2Re\left(-\sqrt{T}\alpha\right)\right]\\
        \nonumber &+\frac{1}{4}\left[2Im\left(i\sqrt{T}\alpha\right)-2Im\left(-i\sqrt{T}\alpha\right)\right].
    \end{align}

    On the other hand, $v$ is Bob's variance, that is, the sum of the modulation variance $2T\alpha^2$ and the variance associated with noise $1+T\xi$.
\end{Observation}

In the same paper, the authors propose a way to extend their results to more general constellations, particularly to QAM. Furthermore, in \cite{Denys2021}, the authors manage not only to define an analogous convex optimization problem but also to find an analytical solution to it. Consequently, there is currently a tool that provides an expression for the optimal parameter \( Z \) (usually denoted by \( Z^* \)). At this point, the Gaussian extremality property ensures that
\begin{equation}
    \chi_{\hat{\rho}_{AB}^{*G}}(Y;E)\geq \sup_{\hat{\rho}_{AB}}\chi_{\rho_{AB}}(Y;E),
\end{equation}
where $\hat{\rho}^{*G}_{AB}$ denotes the Gaussian state with CM 
\begin{equation}
    \begin{pmatrix}
    V\mathbb{I} & Z^*\sigma_z\\
    Z^*\sigma_z & v\mathbb{I}
\end{pmatrix}.
\end{equation}

\begin{figure}[htbp]  
    \centering  
    \begin{tabular}{c}
        \includegraphics[width=0.45\textwidth]{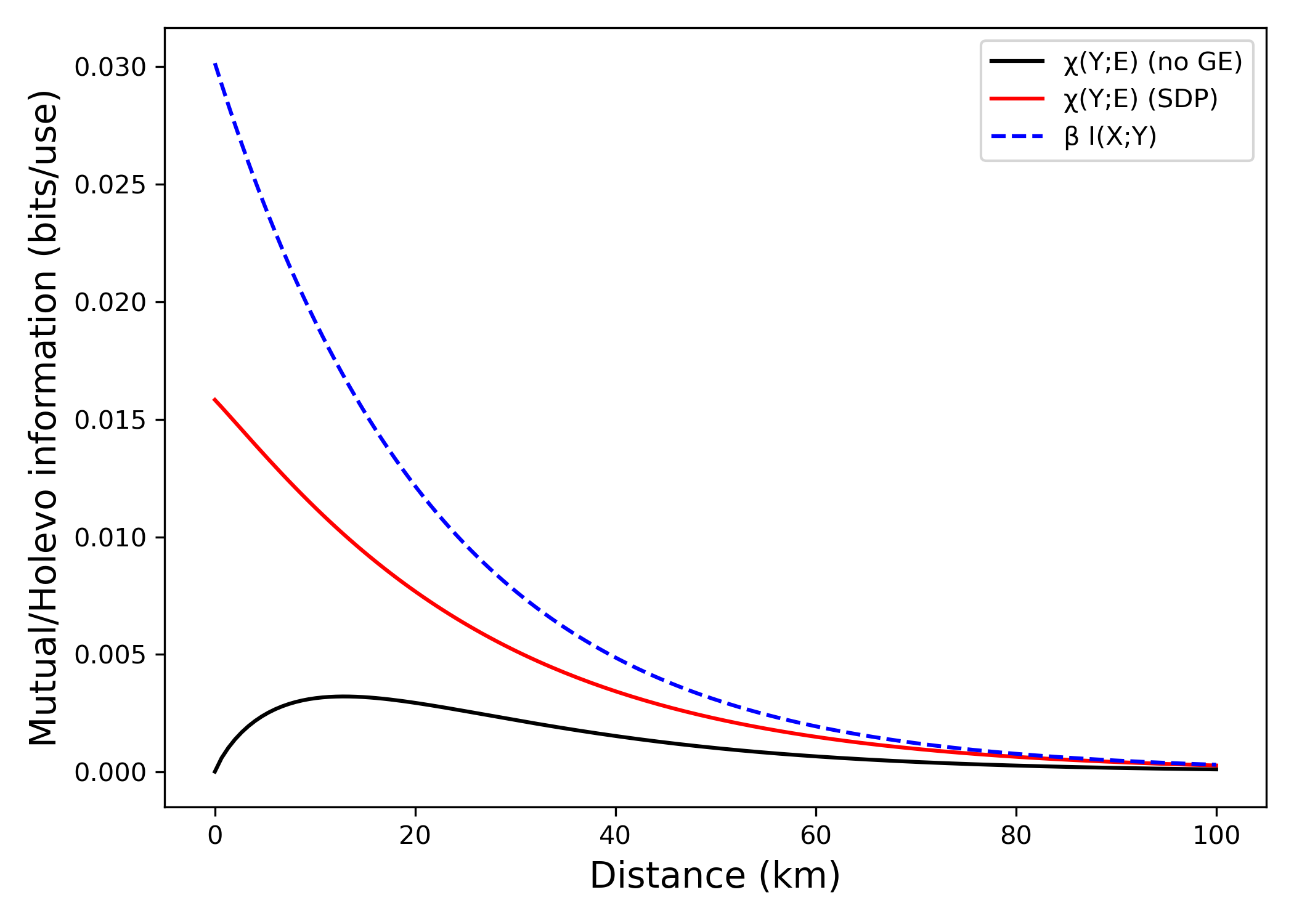} \\
        
    \end{tabular}

    \vspace{1.5ex}

    \begin{tabular}{c}    
        \includegraphics[width=0.45\textwidth]{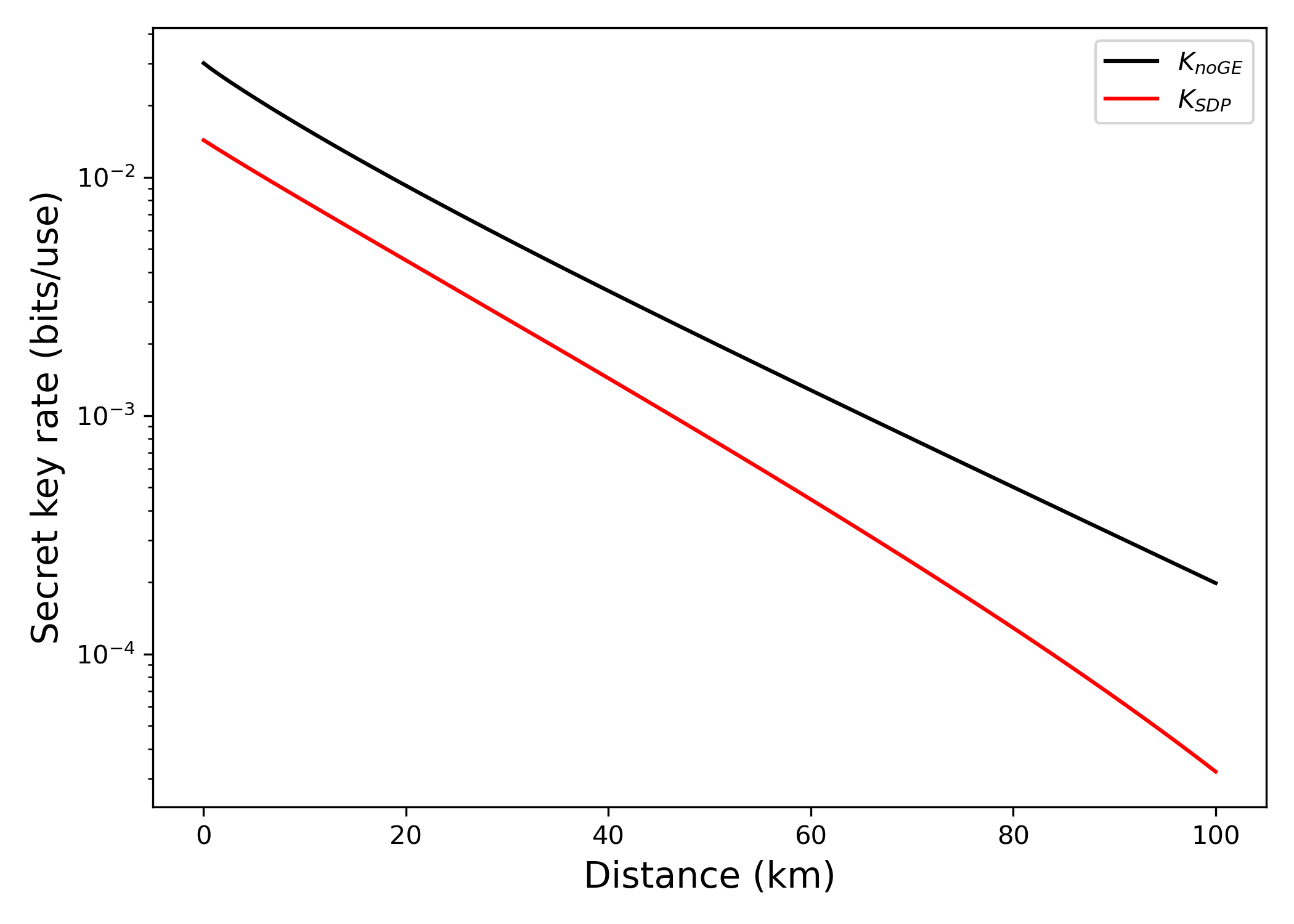} \\
    \end{tabular}  
    \caption{The top figure shows the Holevo information with and without the Gaussian extremality property for the red and black curves, respectively. The red curve is obtained using semidefinite programming (SDP), following \cite{Denys2021}, while the black curve is based on direct evaluation (Eq.\eqref{holevonoGE}). The blue dashed curve shows the mutual information assuming a reconciliation efficiency of $\beta = 0.95$. The bottom figure displays the corresponding secret key rates (Eq.\eqref{eq:DW}) for the black and red Holevo curves in the top figure. This scenario considers a pure-loss channel and BPSK modulation with coherent state amplitude $\alpha = 0.15$, under the assumptions of homodyne detection and reverse reconciliation.}  
    \label{fig:grafica}  
\end{figure}

As should be clear by now, the Gaussian extremality property is extremely useful, as it allows one to reduce the evaluation of entropy quantities, which would otherwise be difficult to compute. However, the price to pay is also significant: this approach tends to overestimate Eve's knowledge about the information exchanged between Alice and Bob (see Fig.~\ref{fig:grafica}).

\subsubsection{Parameter estimation for QPSK}\label{Parameter estimation} 

Now, we present an example of parameter estimation. As demonstrated throughout this section, it is a crucial step in the security proof for DM. Although it is not regarded as essential for an initial approach to understanding the available literature, it is commonly used without a detailed explanation of the process.  

Although the channel is typically characterized by its transmittance $T$ and the excess noise $\xi$, which are inferred during the post-processing stage, these parameters can also be evaluated from the variance parameter $v$ and covariance parameter $c$, since both are functions of $T$ and $\xi$, i.e., $v = v(T, \xi)$ and $c = c(T, \xi)$.  In the case of QPSK modulation presented in \cite{ghorai2019asymptotic}, we have (Observation \ref{observation1})
\begin{equation}\label{eq:estimation}
    v(T, \xi) = 1 + 2 T \alpha^2 + T \xi \hspace{0.3cm} \text{and} \hspace{0.3cm} c(T, \xi) = 2 \sqrt{T} \alpha.
\end{equation}

Now, these parameters can be calculated through estimators of the variance and covariance of the classical variables from Alice and Bob, which carry the values associated with the sent states and the values obtained after measuring the received states.  
Then, after $N$ rounds of the protocol, and assuming the heterodyne detection, Alice has a sequence of values $X = (x_1, x_2, \ldots, x_N)$ where  
$x_i=(x_{i1},x_{i2}) \in \{ (1,0), (-1,0), (0,1), (0,-1) \}$\footnote{In this case, the vectors $(1,0), (-1,0), (0,1), (0,-1)$ represent the choice among the states that Alice sends through the channel $\ket{\alpha}, \ket{-\alpha}, \ket{i \alpha}, \ket{-i \alpha}$, respectively.} (see Observation \ref{observation1}),  
and Bob has a sequence $Y = (y_1, y_2, \ldots, y_N)$ where each $y_i$ is the measurement outcome on the received state, i.e., $y_i=(y_{i1},y_{i2}) \in \mathbb{R}^2$. Thus,
\begin{eqnarray}
    c = \frac{1}{2n}\sum_{i=1}^N \sum_{j=1}^2 x_{ij} y_{ij}.
\end{eqnarray}
We can then estimate $\sqrt{T}$ using Eq.\eqref{eq:estimation}, obtaining
\begin{equation}
    \sqrt{T} = \frac{c}{2 \alpha}.
\end{equation}

Additionally, we know that a good estimator for the variance of a variable is obtained from the sample variance, so an estimator for $v$ is given by:
\begin{eqnarray}
    v = \frac{1}{2N} \left( \sum_{i=1}^N \sum_{j=1}^2 y_{ij}^2 \right).
\end{eqnarray}

Therefore, from Eq.\eqref{eq:estimation}, it follows that
\begin{equation}
    \xi = \frac{v - 1 - 2 T \alpha^2}{T}.
\end{equation}

For general constellations, the variance and covariance terms $v$ and $c$ must be modified to match the protocol description in \cite{Denys2021}.

Additionally, this is not the only way to perform parameter estimation. Starting from the relation in Eq.\eqref{eq:channel_model}, the values of $\sqrt{T}$ and the variance of $\Xi$ can be estimated using linear regression techniques \cite{montgomery2021introduction}, as shown in Sec.~\ref{sec6}.

At this point, an observation about the parameter estimation process and its role in security analysis is necessary. This process focuses on determining certain parameters from experimentally accessible quantities measured by Alice and Bob. For this purpose, they disclose all information about a randomly chosen set of rounds and then process the data by computing quantities such as variance and covariance (as shown previously) or, for example, the first and second moments of the $\hat{q}$ and $\hat{p}$ quadratures conditioned on each of the states that Alice sends \cite{Lin2019}.

In the case of Gaussian modulation, where the optimal attack is known, Alice and Bob can completely determine the state $\hat{\rho}_{AB}$ by finding the estimated values of $T$ and $\xi$ and recovering the covariance matrix of the state of Eq.\eqref{eq:CMchannel}. In the case of discrete modulation, where Eve's optimal attack is unknown, these quantities or parameters allow Alice and Bob to constrain the state $\hat{\rho}_{AB}$, and the key rate is calculated by imposing these constraints on the optimization problems formulated to compute quantities associated with the amount of information Eve possesses (as in the example of Eq.\eqref{SDP} or, analogously, the optimization problem proposed in \cite{Denys2021explicitasymptotic}, or with a different approach in \cite{Lin2019}). If the key rate calculation shows that no secret keys can be generated, then the protocol is aborted.

\subsubsection{Beyond Gaussian extremality property} 

The possibility of reformulating the Devetak–Winter rate using results from information theory (including those initially proven in the finite-dimensional setting and later extended to CV-QKD \cite{Winick2018reliablenumerical, Coles2016}), combined with the use of convex optimization methods, opened the door to new security proofs for protocols with larger constellations, without relying on the Gaussian extremality property.

In this context, the work of Lin et al. \cite{Lin2019} builds on the approach developed in earlier studies \cite{coles2011, coles2012}, employing semidefinite programming to derive a lower bound on the key rate. Lin et al.'s work adopts a more complex framework, which goes beyond the scope of our current objective.

As our intention here is to present important considerations for newcomers to the field of CV-QKD, we focus this section on elucidating a few selected concepts that we ourselves found particularly challenging when first approaching the topic of security proofs.

Nonetheless, we emphasize that the work of Lin et al. is highly relevant and deserves to be revisited, as, just like the work of Ghorai et al. \cite{ghorai2019asymptotic}, it contributes significantly to clarifying and advancing a broader body of research inspired by or closely related to these developments.

This new approach, although computationally more demanding than methods based on Gaussian extremality, demonstrates significantly higher key rates, especially in moderate to high loss regimes \cite{Usenko2025,Lin2019}.

Up to this point, we have focused on detailing key aspects of classical CV-QKD protocols based on the prepare-and-measure paradigm, as well as specific features of security proofs in the asymptotic regime. In the following section, we provide a brief introduction to protocols that explore alternative assumptions beyond the PM model and the asymptotic regime, with the aim of sparking readers’ curiosity about topics of significant current interest.  

\section{Beyond PM protocols and asymptotic security}

\subsection{MDI-CV-QKD protocols}
\label{sec:MDI}

The MDI-QKD protocol represents a crucial advance in secure communications, offering robust protection against detector-side vulnerabilities that affect traditional QKD systems \cite{Pirandola2015, Fletcher2025}. By delegating all quantum measurements to an untrusted central relay (Charlie), the MDI protocol removes security loopholes associated with detectors, which are common targets for side-channel attacks \cite{Braunstein2012, Lo2012, Gerhardt2011}. MDI-CV-QKD combines the advantages of CV-QKD and MDI-QKD to enable high-rate secure key distribution over metropolitan distances \cite{Fletcher2025}. The first protocol, which provided both the theoretical framework and an experimental demonstration, was presented in \cite{Pirandola2013,Pirandola2015}, and GM variants were subsequently proposed by Ma \textit{et al.} \cite{Ma2014} and Li \textit{et al.} \cite{Li2014}. Within the context of coherent attacks, Ottaviani \textit{et al.} \cite{Ottaviani_2015} analyzed eavesdropping strategies on quantum links in the symmetric model, demonstrating the superiority of two-mode attacks over single-mode collective attacks based on independent entanglement-cloning strategies. In the context of collective attacks, Zhang \textit{et al.} \cite{Zhang2014} introduced a GM MDI-CV-QKD protocol employing squeezed states, showing higher secret key rates. Moreover, Chen \textit{et al.} \cite{Chen2018} extended the analysis to coherent attacks and provided a composable security assessment through the application of entropic uncertainty relations.

A variety of techniques have been proposed to improve protocol performance, including non-Gaussian operations (photon subtraction \cite{Zhao2018,Ma_2018,Yu_2022,Djordjevic_2019}, quantum catalysis \cite{Ye2020,BilalKhan:23,Kumar:25}, quantum scissors \cite{Jafari:2022svx}), amplifiers (noiseless linear \cite{Zhang_2015}, phase sensitive \cite{Wang2019}), postselection \cite{Wilkinson2020}, multi-mode Gaussian modulation \cite{Ding2021}, and free-space implementations \cite{Ghalaii2023}. Alternative encoding strategies have also been studied with thermal state encodings \cite{Bai_2019}, unidimensional modulation \cite{Bai:2019}, and discrete or dual-phase replacements for Gaussian modulation \cite{Ma2019,Liao:18}. Recent efforts concentrate on scaling MDI-CV-QKD to multi-user networks with untrusted intermediate nodes \cite{Wu_2016,Ottaviani2019,Fletcher2022} and on assessing realistic performance and post-processing within the composable finite-size security framework \cite{Mountogiannakis2022}. 

In this section, we provide a general discussion on MDI-CV-QKD protocols. A more detailed analysis of the main features and the most recent advancements in MDI-CV-QKD can be found in Ref.\cite{Fletcher2025}.

\subsubsection{Gaussian MDI-CV-QKD protocol}

The framework of the MDI-CV-QKD protocol proposed by \cite{Pirandola2015} is structured as follows: Alice and Bob independently encode their information into the quadratures of coherent states by applying Gaussian modulations to their amplitudes. Specifically, Alice initiates the protocol by preparing the mode $A$ in a coherent state \(|\alpha\rangle_A:=|\alpha\rangle\), with a Gaussian distribution with zero mean and variance (see Sec.~\ref{cvqkddescription}). Likewise, Bob prepares the mode $B$ in a coherent state \(|\alpha\rangle_B:=|\beta\rangle\), which is drawn from the same Gaussian distribution. The coherent states are transmitted through a potentially insecure quantum channel to a central relay, which performs a CV Bell measurement, which is a joint measurement that projects onto EPR eigenstates \(\hat{q}_-\) and \(\hat{p}_+\). Practically, this is implemented by interfering Alice’s and Bob’s modes on a balanced beam splitter, followed by two homodyne measurements on the output ports. 

In the entanglement-based picture of the protocol, the same quantum correlations are reproduced by having Alice and Bob each generate TMSVS (see Eq.(\ref{tmsvs})), with the covariance matrix given by Eq.(\ref{tmsvscm}),  and sending one mode of each state to the central relay while keeping the other modes locally. On the modes kept locally by Alice and Bob, heterodyne measurements are performed, effectively projecting coherent states onto the modes sent to the central relay. It should be noted that, due to the commutativity of local measurements, Alice’s and Bob’s heterodyne detections can be postponed until after Charlie’s measurements \cite{Pirandola2015}. In this way, the protocol can be interpreted as an entanglement swapping procedure \cite{Pirandola_2006}, where Alice’s and Bob’s local modes become entangled as a result of Charlie’s measurements.  

The modes sent by Alice and Bob to the central relay propagate through the untrusted quantum link channels \(L_{AC}\) and \(L_{BC}\) (see Fig.~\ref{fig:fig0}), accessible to Eve before reaching the relay station operated by Charlie, who may potentially be under Eve's control. It should be emphasized that the links \(L_{AC}\) and \(L_{BC}\) can exhibit equivalent distances (symmetric model) or different distances (asymmetric model), resulting in distinct secret key rates \cite{Pirandola2015}. 

\begin{figure}[h]  
    \centering  
    \includegraphics[width=0.48\textwidth]{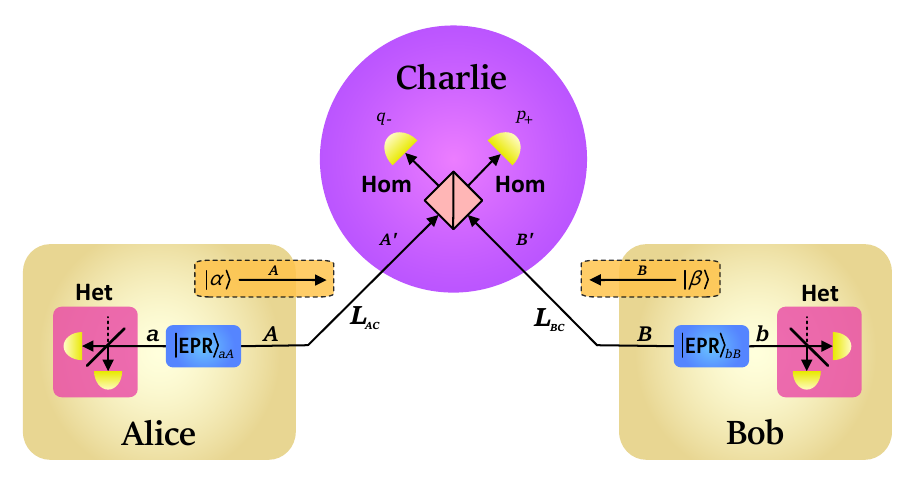}  
    \caption{PM and EB pictures of the MDI-CV-QKD protocol. In the PM representation (dashed box), Alice and Bob transmit the Gaussian modulated coherent states \(|\alpha\rangle \) and \(|\beta\rangle\), respectively, to the central relay via the links \(L_{AC}\) and \(L_{BC}\), where signal modes \(A\) and \(B\) are transformed into \(A'\) and  \(B'\) by Gaussian channels before reaching the central relay (see Sec. \ref{quantumchannel}). At the central relay, Charlie receives modes \(A'\) and \(B'\), mixes them on a balanced beam splitter, and performs two conjugate homodyne detections, yielding the real values \(q_{-}\) and \(p_{+}\), which are combined to extract the complex variable \(\gamma\), communicated over a public and authenticated classical channel. In the EB picture, the preparation carried out by Alice and Bob consists of generating a two-mode EPR state and performing a heterodyne detection on one of the modes, which projects the other mode onto a coherent state. This coherent state is then transmitted to the central relay, in analogy with the PM model.}  
    \label{fig:fig5}  
\end{figure}

Upon arrival at the central relay, Charlie implements a CV Bell detection: the incoming modes from Alice and Bob interfere on a balanced beam splitter, and Charlie performs two homodyne measurements (one on each output port) to obtain the quadratures $\hat{q}_{-}=(\hat{q}_A-\hat{q}_B)/\sqrt{2}$ and $\hat{p}_{+}=(\hat{p}_A+\hat{p}_B)/\sqrt{2}$. He then publicly announces the complex outcome \(\gamma:= (q_{-}+ip_{+})/\sqrt{2}\), which (in the ideal, lossless or noiseless case for inputs \(|\alpha\rangle\) and \(|\beta\rangle\)) equals \(\alpha - \beta^*\). Alice and Bob use this public value \(\gamma\) to correlate their strings and during the parameter estimation process \cite{Pirandola2015}. In Fig.~\ref{fig:fig5}, we illustrate the schematic of the MDI-CV-QKD protocol in both the PM and the EB representations.

After receiving the measurement results publicly announced by Charlie, Alice and Bob employ post-processing to correlate their raw data \cite{Pirandola2015}, thereby generating the mutual information required for key extraction, which was absent prior to the relay’s measurement. In contrast, Eve, having access only to Charlie’s outcomes, acquires no direct information about Alice’s or Bob’s individual variables and is therefore forced to attack the communication links. From these correlated data, Alice and Bob perform parameter estimation, error correction, and privacy amplification, ultimately obtaining the shared secret key (see Sec. \ref{cvqkddescription}).

For the security analysis in Gaussian symmetric MDI-CV-QKD, we adopt the standard worst-case assumption that all channel losses and thermal noise are attributed to Eve, who is therefore modeled as performing the Gaussian attack on both links, \(L_{AC}\) and \(L_{BC}\). In this attack, Alice's and Bob's modes are combined with Eve's auxiliary modes, \(E_1\) and \(E_2\), through two beam splitters with transmissivities \(T_A\) and \(T_B\), respectively (with the symmetric case given by \(T_A=T_B=T\)). The reduced state held by Eve is a zero-mean Gaussian state that incorporates the correlations among her modes, with a covariance matrix in symmetric normal form, is given by \cite{Pirandola_2013}:
\begin{equation}\label{eq: EveCM}
\Sigma_{E_1E_2} = \begin{pmatrix}
    \omega\mathbb{I} & G\\
    G & \omega\mathbb{I}
\end{pmatrix},
\end{equation}
where \(G=\text{diag}(g,g')\) encodes the quantum correlations between Eve’s auxiliary systems, \( \omega\geq1\) denotes the thermal noise variance \(\omega_A=\omega_B\) (where \( \omega=1\) recovers the pure-loss case) \cite{Pirandola2015}, the parameters \(g\), \(g'\) and \(\omega\) satisfy a set of bona-fide conditions \cite{Pirandola_2009}, which ensure consistency with the uncertainty principle. 

The optimal attack allowed for Eve corresponds to the case where  \(g=\sqrt{\omega^2 - 1}=-g'\), this regime is such that the quantum correlations maximize entanglement precisely in opposition to those established by CV Bell measurement performed at the central relay \cite{Ottaviani_2015, Pirandola2015}. In this case, since the variable obtained after the relay is \(\gamma\approx\sqrt{T}\left(\alpha-\beta^{*}\right)\) \cite{Ottaviani_2015}, without loss of generality, we can consider Alice as the encoder of the information (from her heterodyne detection) and Bob as the decoder, meaning that he post-processes his variable \(\beta\) to infer Alice’s variable \(\alpha\) \cite{Pirandola2015}. In the limit of large variances, the mutual information between Alice and Bob is given by \(I_{AB}=\log\left(\frac{V}{\xi}\right)\), where the noise parameter \(\xi:=\xi(T,\omega,g,g')=\xi_{pure-loss}+\varepsilon\), with \(\xi_{pure-loss}\) describing the pure-loss noise and \(\varepsilon\) representing the excess noise, can be determined from the probability distribution associated with the joint statistics of \(\alpha\), \(\beta\), and \(\gamma\) that must be recovered \(p(\alpha,\beta,\gamma)\) \cite{Pirandola2015}. 

On the other hand, Eve's Holevo information is obtained from \(\chi_E =S(\hat{\rho}_{ab|\gamma})-S(\hat{\rho}_{b|\gamma\tilde{\alpha}})\), where \(S(\hat{\rho}_{ab|\gamma})\) denotes the entropy of the joint state of Alice and Bob conditioned on Charlie’s measurement outcome, representing the remaining quantum correlations available before conditioning on Alice’s classical information, and \(S(\hat{\rho}_{b|\gamma\tilde{\alpha}})\) represents the entropy of Bob’s state after accounting for Alice’s classical variable (in addition to \(\gamma\)). Here \(\tilde{\alpha}\) denotes the complex-valued variable obtained in the EB representation, corresponding to the outcome of Alice’s heterodyne measurement on one mode of the EPR state. Analogously discussed in Sec.~\ref{subsecCollect}, assuming unit reconciliation efficiency and an infinite number of protocol rounds, the secret key rate shared by Alice and Bob, corresponding to the average number of secret bits per use of the relay is given by \(K = I_{AB}-\chi_E\) \cite{Pirandola2015, Ottaviani_2015}, where both the mutual information between Alice and Bob and Eve’s Holevo information are conditioned on the variable  \(\gamma\). For the limit of large variances (\(V_{Mod}>>1\)) and the symmetric regime of the protocol considered here, the secret key rate is given by \cite{Pirandola2015}:
\begin{equation}
   K_{Sym} = \log\left(\frac{16}{e^{2}\xi\left(\xi-4\right)}\right) + g\left(\frac{\xi}{2}-1\right),
\end{equation}
with \(g(x)\) defined in Eq.(\ref{eq:g function_}).

For further theoretical details on the security analysis of the Gaussian MDI-CV-QKD protocol, see the supplementary material of Ref.\cite{Pirandola2015}.

\subsubsection{Discrete modulation MDI-CV-QKD protocol}

In 2019, Ma \textit{et al.} \cite{Ma2019} proposed an MDI-CV-QKD protocol based on four-state discrete modulation, using non-orthogonal coherent states to encode bits. This approach enables longer transmission distances in the low-SNR regime \cite{Ma2019} and simplifies implementation compared to Gaussian modulation. In such DM MDI-CV-QKD schemes, Alice and Bob independently prepare their states and apply their respective modulation operations. An example of DM is the QPSK scheme, described by four coherent states that are phase-shifted by \(\pi/2\) relative to each other \cite{leverrier2009unconditional}:
\begin{equation}
\left\{ \ket{\alpha e^{i\pi/4}},\ket{\alpha e^{3i\pi/4}},\ket{\alpha e^{-3i\pi/4}},\ket{\alpha e^{-i\pi/4}}\right\}.
\end{equation}

In the prepare-and-measure representation, the states sent by Alice (Bob) to the central relay through the untrusted channel are described as a statistical mixture of coherent states  (see Eq.(\ref{eq:tau_state}))
\cite{leverrier2009unconditional, Ma2019}:
\begin{align}
\hat{\rho}^{A(B)}_4 = \frac{1}{4} \sum_{j=0}^3 \ket{\alpha_j}\bra{\alpha_j}
\end{align}
which can be conveniently expressed in terms of the following diagonalization \cite{Denys2021}:
\begin{align}
\hat{\rho}^{A(B)}_4 =\sum_{j=0}^3 p^{A(B)}_{j}\ket{\phi_j}\bra{\phi_j}_{A(B)}, 
\end{align}
with
\begin{equation}
\begin{cases}
p^{A(B)}_{0,2} &= \frac{1}{2e^{\alpha_{A(B)}^2}} \left[ \cosh\left(\alpha_{A(B)}^2\right) \pm \cos\left(\alpha_{A(B)}^2\right) \right] \\
p^{A(B)}_{1,3} &= \frac{1}{2e^{\alpha_{A(B)}^2}} \left[ \sinh\left(\alpha_{A(B)}^2\right) \pm \sin\left(\alpha_{A(B)}^2\right) \right]
\end{cases}
\end{equation}
 and
\begin{align}
\ket{\phi_j}_{A(B)} =& \frac{e^{-\alpha^2/2}}{\sqrt{p^{A}_j}} \sum_{n=0}^\infty (-1)^n \frac{\alpha^{4n+j}}{\sqrt{(4n+j)!}} \ket{4n+j},
\end{align}
with \(j\in \{0,1,2,3\}\).
In the EB picture of the protocol, in which Alice and Bob independently prepare the non-Gaussian states \cite{Denys2021, Ghorai2019}:
\begin{align}
\ket{\Psi_4}_{aA(bB)} =& \sum_{j=0}^3 \sqrt{p^{A(B)}_j} \ket{\bar{\phi}_j}_{a(b)}\otimes\ket{\phi_j}_{A(B)} \nonumber \\=& \frac{1}{2} \sum_{j=0}^3 \ket{\psi_j}_{a(b)} \otimes \ket{\alpha_j}_{A(B)},
\end{align}
where the states \(|\alpha_{j}\rangle_{A}\) and \(|\alpha_{j}\rangle_{B}\) are coherent states generated by Alice and Bob, respectively. The non-Gaussian states prepared by Alice and Bob possess the same structure as those introduced in Sec. (\ref{subsec:DM}), in the same manner, the symmetric form for the covariance matrices, presented in Eq.(\ref{eq: sym CM}), of Alice and Bob can be obtained from the symmetrization arguments discussed in Sec. (\ref{subsec:DM}). Thus, it is worth noting that in the specific context presented in \cite{Ma2019}, the modulation variance of Bob was assumed to be equivalent to that of Alice, as well as the covariance matrix associated with the state produced by Bob. Consequently, in each round, Alice and Bob effectively send one of their four QPSK-modulated coherent states to the central relay. In the relay, Charlie performs the CV Bell measurement and publicly announces the results. Alice retains the original sign of her quadratures, while Bob applies a displacement operation conditioned on Charlie’s announced results. Thus, Alice and Bob are able to perform post-processing by implementing parameter estimation, information reconciliation, and privacy amplification, as in conventional CV-QKD protocols, to obtain a fully correlated and secret key sequence of bit strings. The security analysis of this protocol is carried out in the asymptotic limit against collective attacks, with the use of decoy states.

As mentioned in the introduction, in this section, we provide only a brief discussion of the architecture of MDI-CV-QKD protocols while indicating the main references that present a more in-depth treatment of these protocols.


\subsection{Finite-size security}
\label{sec6}

Finite-size effects significantly impact the security of CV-QKD protocols by introducing statistical fluctuations that must be carefully accounted for in practical implementations \cite{Leverrier2010, Pascual2025, Furrer100502}. Since only a finite number of signals can be exchanged, incorporating finite-size corrections becomes essential to ensure the security of the generated key, even in the presence of an adversary with unbounded quantum capabilities \cite{Pirandola043014, Jain2022}. This section addresses two main challenges arising from finite data size: parameter estimation and composable finite-size security analysis. The theoretical aspects of both challenges are discussed here, while practical implementations and examples can be found in refs. \cite{Leverrier2010, Jain2022, Furrer100502}.

\subsubsection{Parameter estimation}

Since QKD assumes the quantum channel is fully under Eve's control, Bob and Alice cannot trust the channel parameters \cite{Gisin145, Scarani2009}. Thus, a fundamental procedure in CV-QKD is the estimation of the channel parameters, such as the transmittance $T$ and the excess noise $\xi$ \cite{Leverrier2010}. In practice, these are not the only parameters that must be estimated in a real implementation, such that one also needs to account for Alice’s modulation variance $V_A$ and the quantum efficiency of the detectors $\eta$ in scenarios where Bob’s detection is calibrated. However, it is reasonable to assume that these parameters are relatively well known compared to $T$ and $\xi$, where the latter has a drastic impact for long distances \cite{leverrier2009, Wilkinson2020, Zhang2024, Weedbrook2004}.

In general, Gaussian channels can be described by the normal linear model in Eq.\eqref{eq:channel_model} without loss of generality, where $t = \sqrt{T}$ and the random variables have the distribution $X \sim \mathcal{N}(0, V_A)$ and $\Xi \sim (0, \sigma^2)$, with $\sigma^2 = \mu + T\xi$ being the noise variance \cite{Pirandola043014}. In a generic CV-QKD protocol, Alice modulates and sends $N$ signals through the quantum channel, where every signal represents a quadrature measurement \cite{Leverrier2010}. In principle, the parameter estimation can be done by one of the authenticated parts, as long as the part responsible for this process has access to $\{X\}_{m}$ and $\{Y\}_{m}$ signals, where $m = N-n$ \cite{Pirandola043014}. Thus, $n$ raw signals are left to generate the secret key.

It is well-established in the literature that the Maximum Likelihood Estimation (MLE) method offers robust security guarantees for CV-QKD protocols~\cite{ghalaii2020long, Papanastasiou042332, Pirandola023321, Thearle042343}. This is substantiated by its foundational role in the first security proof against Gaussian collective attacks in the finite-size regime~\cite{Leverrier2010}, a result that established the standard framework to take into account finite-size parameter estimation effects under various scenarios ~\cite{Furrer100502, Jain2022, Papanastasiou042332, Mountogiannakis2022, Pascual2025}. For the relevant channel parameters, the corresponding estimators derived from these linear models are given by
\begin{equation} \label{eq:MLE}
    \hat t = \sum_i^m \frac{y_i x_i}{x_i ^2} \, \quad \text{and} \, \quad \hat \sigma^2 = \frac{1}{m}\sum_i^m (y_i - \hat tx_i)^2 \,,
\end{equation}
with distributions
\begin{equation}
    \hat{t} \sim \mathcal{N}\!\left(t, \frac{\sigma^{2}}{\sum_{i=1}^{m} x_i^{2}}\right)
\quad \text{and} \quad
\frac{m \hat{\sigma}^{2}}{\sigma^{2}} \sim \chi^{2}(m-1)\,,
\end{equation}
where the $\chi^2$ converges to a normal distribution in the regime of large samples \cite{monfort1982cours}. 

Since $\hat{t}$ and $\hat{\sigma}$ are unbiased\footnote{One can consider that the given variance estimator is unbiased for large sample size \cite{casella2024statistical}.} estimators for the transmittance and variance,  $\mathbb{E}[\hat t] = t$ and $\mathbb{E}[\hat \sigma^2] = \sigma^2$, with the corresponding variances decreasing with the sample size. Nevertheless, for a finite sample, the variances should always be non-zero, and the estimated values will differ slightly from the true parameter values. In this case, different interval cases for the estimators need to be considered: 

\begin{enumerate}
    \item \textbf{Optimistic-case}: $\hat t = t$ or $\hat \sigma^2 = \sigma^2$. This is the best possible scenario, where the estimator converges to the real values of the parameters. However, the probability that an estimate converges to the mean of its estimator is very low in the low-sample scenario. In the asymptotic scenario, the optimistic-case is always guaranteed.
    
    \item \textbf{Overestimated-case}: $\hat t > t$  or $\hat \sigma^2< \sigma^2$. This is an insecure scenario for the authenticated parts, where the estimators imply an untrusted value. If one of the authenticated parts uses this estimation, the secret-key rate can be overestimated, which implies that unconditional security is no longer guaranteed.
    
    \item \textbf{Worst-case}: $\hat t < t$ and $\hat \sigma^2 > \sigma^2$. This is the conservative case, where the unconditional security is strongly guaranteed, since one of the authenticated parties always underestimates the secret-key rate. Thus, the challenge is to get close to the \textit{optimistic-case} without ever reaching the \textit{overestimated-case}.
\end{enumerate}

Therefore, there must exist an error parameter $\epsilon_{\mathrm{PE}}$ able to ensure the worst-case confidence intervals for the estimators in Eq.\eqref{eq:MLE}, such that all the points estimated are inside the confidence intervals with probability at least $1 - \epsilon_{PE}/2$. These values are then given by
\begin{equation} \label{eq:t_min}
    t_{\min} \approx \hat t - z_{\epsilon_{PE}/2} \sqrt{\frac{\hat \sigma^2}{m V_A}}
\end{equation}
and 
\begin{equation} \label{eq:sigma_max}
  \sigma^2_{\max} \approx \hat \sigma^2 + z_{\epsilon_{PE}/2} \frac{\hat \sigma^2 \sqrt{2}}{\sqrt{m}} \,,
\end{equation}
where $z_{\epsilon_{PE}/2} = \text{erf}^{-1}(1 - \epsilon_{PE}/2)$ and $\mathrm{erf}(x)$ is the error function. 

In this context, the covariance matrix assuming the probability of failure of MLE is rewritten as
\begin{equation} \label{eq:cov_mat_worst}
    \Sigma{\epsilon_{PE}} =  \begin{pmatrix}
(V_A + 1)\mathbb{I}_2 &  t_{min} Z \sigma_z\ \\
t_{min} Z \sigma_z &  (t_{min}^2V_a + \sigma^2_{max})\mathbb{I}_2 \\
\end{pmatrix} \,,
\end{equation}
which implies the existence of a confidence set $\mathcal{C}_{\epsilon_{PE}}$ such that the covariance matrix $\Sigma_{\epsilon_{PE}}$ belongs to $\mathcal{C}_{\epsilon_{PE}}$ with probability at least $1 - \epsilon_{PE}/2$ \cite{Furrer100502}. Therefore, the Holevo information in Eq.\eqref{eq:HolInf} is bounded by the worst-case confidence intervals, resulting in
\begin{equation}
    k_{\epsilon_{\mathrm{PE}}} \geq \frac{n}{N}(\beta I(X;Y) - \chi_{\epsilon_{\mathrm{PE}}}(Y;E)) \,.
\end{equation}
where the $m$ signals discarded result in a fraction of $n/N$ signals left.


\subsubsection{Composable finite-size key}

The notion of universal composability was formalized by Canetti in his seminal work \cite{Canetti959888}, which introduced a real-world/ideal-world paradigm. This framework comprises an ideal protocol and a real protocol, both of which include communicating parties and an adversary. The key innovation of this model was the introduction of an environment machine, which has access to the outputs of both protocols. The central challenge in cryptography is therefore to quantify the environment's ability to distinguish between the real and ideal worlds \cite{Usenko2025}. In the QKD scenario, composability ensures that the security errors from each step of the protocol combine into an overall security parameter \cite{Pirandola043014}. This guarantees that the secret key can be securely employed in any subsequent cryptographic application \cite{Canetti959888, Scarani200501}.

In this context, a QKD protocol is considered secure when it satisfies two conditions: correctness and secrecy \cite{Tomamichel2012, Usenko2025, Pirandola023321, Renner05}. Correctness means that the final keys produced by Alice, $S_A$, and Bob, $S_B$, must agree ($S_A = S_B$). Secrecy requires that Alice’s key $S_A$ is uniformly random and entirely independent of any quantum system $E$ accessible to an eavesdropper. Both requirements depend on methods that require many samples adopted during post-processing \cite{math12142243}, such that one must have a sufficient amount of data in order to ensure these requirements.  

\subsubsubsection*{Correctness}

The correctness of the protocol depends essentially on the reconciliation information procedure \cite{Zhang2024, Diamanti2015}. The major challenge in this step is to design error-correcting codes for long-distance transmission with a very low signal-to-noise ratio of the optical quantum channel \cite{leverrier2009unconditional, zhang2020, Ruppert2014}. At such low SNR levels, efficient key reconciliation is possible only with low-rate block codes that use very large block sizes \cite{Young905935}. In this context, a key parameter used to quantify how much information can be recovered in this process is the reconciliation efficiency, which can limit the mutual information. 

Following parameter estimation, each block of size $N$ yields $n$ usable signals, which are subsequently processed into a shared secret key through error correction and privacy amplification. For a given block, error correction succeeds with probability $p_{\mathrm{EC}}$, while the complementary failure probability, $\mathrm{FER} = 1 - p_{\mathrm{EC}}$, is referred to as the ``frame error rate" \cite{Milicevic2018}. The success probability $p_{\mathrm{EC}}$ is determined by the SNR ratio, the target reconciliation efficiency $\beta$, and the $\epsilon_{\mathrm{cor}}$-correctness parameter \cite{Pirandola043014}. In this sense, $np_{\mathrm{EC}}$ signals are, on average, sent to the privacy amplification step, resulting in 
\begin{equation}
    k_{\epsilon_{\mathrm{PE}} + \epsilon_{\mathrm{cor}}} \geq \frac{np_{\text{EC}}}{N} (\beta I(X;Y) - \chi_{\epsilon_{\mathrm{PE}}}(X;E)) \,,
\end{equation}
which ensures that the protocol is $\epsilon_{\text{cor}}\text{-correct}$ for $\text{Pr}[S_A \neq S_B] \leq \epsilon_{\text{cor}}$. Note again that, even though $\mathrm{FER} \cdot n$ signals are discarded, this is still negligible in the asymptotic scenario. 

\subsubsubsection*{Secrecy}

The joint state of the classical register $S_A$ and the adversary's quantum system $E$ can be expressed as a classical-quantum state,
\begin{equation}
    \omega_{S_AE} = \sum_{s} |s\rangle\langle s| \otimes \omega_E^s ,
\end{equation}
where $\{\omega_E^s\}_s$ are the quantum states on Eve’s system $E$ conditioned on the classical key value $s$ \cite{Furrer100502}.

To ensure the secrecy of the protocol, the ideal joint state factorizes as $\omega^{\mathrm{id}}_{S_AE} = \tau_{SA} \otimes \sigma_E$, where $\tau_{SA}$ denotes the maximally mixed state over the key space, and $\sigma_E$ is an arbitrary quantum state on Eve’s system $E$ \cite{Renner05, Furrer100502, Tomamichel2012}. Thus, a key distribution protocol is called $\epsilon_s$-secret if the real classical-quantum state $\omega_{S_AE}$ is $\epsilon_s$-close to the ideal state in terms of the trace distance. 
Formally, this is expressed as
\begin{equation}
    \inf_{\sigma_E} \,
    \frac{1}{2} \left\| \omega_{S_AE} - \tau_{SA} \otimes \sigma_E \right\| \leq \epsilon_s ,
\end{equation}
where the infimum ranges over all normalized states $\sigma_E$ on Eve’s system $E$ \cite{Tomamichel2012}.

In post-processing, the secrecy parameter $\epsilon_s$ depends primarily on two factors: a penalty term from the asymptotic equipartition property (AEP) and the privacy amplification step. The first one arises because the smooth min-entropy serves as the operational measure for the extractable secret key length in the finite-size regime; it converges to the von Neumann entropy only asymptotically \cite{Leverrier2010}. Since the security analysis is based on the von Neumann entropy, it is necessary to account for the convergence speed of the smooth min-entropy towards this asymptotic value, which is given by 
\begin{equation}
    4 \log (\sqrt{d} + 2)\sqrt{\frac{1}{n}\log_2 \left (\frac{18}{p_{\mathrm{EC}}^2 \bar \epsilon^4} \right )} \,,
\end{equation}
 where  $\bar \epsilon$ is a smooth parameter and $d$ denotes the number of bits per quadrature used during discretization. The derivation of this penalty can be seen in Ref.\cite{Pirandola043014}, where the non-asymptotic framework for the AEP proposed in Ref.\cite{tomamichel2013frameworknonasymptoticquantuminformation} is adopted.

For privacy amplification, the primary challenge is to distill a uniformly random secret key from the raw data. Since extractors used in this stage can only guarantee this requirement perfectly in the asymptotic limit of infinite samples, a finite failure probability $\epsilon_h$ must be accounted for in any practical implementation \cite{PhysRevA.87.062327}. By employing the operational interpretation of the smooth min-entropy, as established in Ref.\cite{Konig5208530}, it can be shown that
\begin{equation}
    \frac{2}{n}\log_2(1/\epsilon_h) \,.
\end{equation}

In general, this value is accounted for the parameter $\Delta(n)$, which is subtracted from the secret-key rate and can be written as:
\begin{equation}
    \Delta(n) = 4 \log (\sqrt{d} + 2)\sqrt{\frac{1}{n}\log_2 \left (\frac{18}{p_{\mathrm{EC}}^2 \bar \epsilon^4} \right )} 
     + \frac{2}{n}\log_2(1/\epsilon_h) \,,
\end{equation}
where $\epsilon_{\mathrm{sec}} = \epsilon_h + \bar \epsilon$. 

Finally, by incorporating all finite-size effects, the achievable secret-key rate is lower bounded by
\begin{equation} \label{eq:skr_finite}
   k_{\epsilon} \geq \frac{n \, p_{\mathrm{EC}}}{N} 
   \left( \beta I(X;Y) - \chi_{\epsilon_{\mathrm{PE}}}(X;E) - \Delta(n) \right) ,
\end{equation}
where the total security parameter is given by
\begin{equation}
    \epsilon = \epsilon_{\mathrm{PE}} + \epsilon_{\mathrm{cor}} + \epsilon_{\mathrm{sec}} \,.
\end{equation}

In particular, the condition \eqref{eq:skr_finite} ensures that the protocol generates an  $\epsilon$-secure key against Gaussian collective attacks, i.e., the distance between the real and the 
ideal key is bounded by $\epsilon$ \cite{Leverrier2010}. As a consequence, the actual secret-key rate satisfies 
\begin{equation}
    k \geq k_{\epsilon},
\end{equation}
with equality in the asymptotic regime.

\section{Final remarks}
\label{secconclusion}

The field of CV-QKD continues to evolve rapidly, with ongoing developments addressing both fundamental and practical challenges. This review has focused primarily on the asymptotic security of PM protocols with coherent states and reverse reconciliation under collective attacks. Several important and more advanced topics have been omitted or only briefly addressed, including the transition to finite-size security analysis, the development of composable security frameworks, and the mitigation of side-channel vulnerabilities—all critical for practical implementations. Furthermore, the integration of CV-QKD with existing telecommunication infrastructure and the development of chip-scale devices promise to make quantum-secure communications more accessible and cost-effective.

For Brazil's emerging quantum communication community, these developments present significant opportunities and challenges. We hope this review provides newcomers with a solid foundation in CV-QKD theory, offering a guided introduction to the key topics we consider most essential for entering the field and clarifying areas where accessible and comprehensive references are scarce. To facilitate further exploration, we have included an extensive bibliography, providing readers with resources to deepen their understanding of any topics that merit additional study.

\begin{acknowledgments}
We thank Leonardo Justino Pereira for carefully reading the manuscript and for insightful comments and discussions. This work was partially funded by the project ``Receptores não-convencionais em CV-QKD'' supported by QuIIN - Quantum Industrial Innovation, EMBRAPII CIMATEC Competence Center in Quantum Technologies, with financial resources from the PPI IoT/Manufatura 4.0 of the MCTI grant number 053/2023, signed with EMBRAPII. MAD thanks financing from the European Union (HORIZON-MSCA-2023 Postdoctoral Fellowship, 101153602 - COCoVaQ).
\end{acknowledgments}

\section*{Declarations}

\textbf{Author contributions:} Maron F. Anka wrote Secs. \ref{sub:cv_systems}, \ref{sec:V}, \ref{subsec6} and prepared figures 1, 5, 6, and 7; John A. M. Rodríguez wrote Sec. \ref{sec5} and prepared figure 4; Douglas F. Pinto wrote Sec.\ref{sec:MDI} and prepared figure 8; Lucas Q. Galvão wrote Sec.\ref{sec6}; Micael A. Dias wrote Sec.\ref{sub:information_theory} and prepared figures 2 and 3; and  Alexandre B. Tacla wrote Secs. \ref{introduction} and \ref{secconclusion} and reviewed the manuscript. All authors contributed to the review of the manuscript.

\textbf{Founding:} This work was partially funded by the project ``Receptores não-convencionais em CV-QKD'' supported by QuIIN - Quantum Industrial Innovation, EMBRAPII CIMATEC Competence Center in Quantum Technologies, with financial resources from the PPI IoT/Manufatura 4.0 of the MCTI grant number 053/2023, signed with EMBRAPII and the European Union (HORIZON-MSCA-2023 Postdoctoral Fellowship, 101153602 - COCoVaQ).

\textbf{Conflict of interest:} The authors declare no competing interests.

\textbf{Data availability:} Not applicable.

\textbf{Code availability:} Not applicable.

\textbf{Materials availability:} Not applicable.

\textbf{Ethics approval and consent to participate:} Not applicable.



\bibliographystyle{unsrt}
\bibliography{bibi}

\end{document}